\documentclass[fleqn,usenatbib]{mnras}

\usepackage[T1]{fontenc}
\usepackage[dvipsnames]{xcolor}
\usepackage{mathtools}
\usepackage{amsmath}
\usepackage{amssymb}
\DeclareRobustCommand{\VAN}[3]{#2}
\let\VANthebibliography\thebibliography
\def\thebibliography{\DeclareRobustCommand{\VAN}[3]{##3}\VANthebibliography}

\usepackage{graphicx}	
\usepackage{amsmath}	
\usepackage{amssymb}	
\usepackage{xcolor}
\usepackage{caption}
\usepackage{subcaption}
\usepackage{newtxtext,newtxmath}

\newcommand{\civ}{\ion{C}{IV}}
\newcommand{\nciv}{N_{\civ}} 
\newcommand{\zciv}{z_{\civ}}
\newcommand{\sciv}{\sigma_{\civ}}
\newcommand{\kms}{km\,s$^{-1}$} 
\newcommand{\zqso}{z_{\textrm{QSO}}}

\newcommand{\dd}{\textrm{d}}
\newcommand{\Data}{\mathcal{D}}
\newcommand{\model}{{\rm M}}
\renewcommand{\doi}[1]{\href{http://dx.doi.org/#1}{doi:#1}}
\newcommand{\arxiv}[1]{\href{http://arxiv.org/#1}{arxiv:#1}}
\definecolor{flatirons}{HTML}{8B2131}
\definecolor{rez}{HTML}{006400}

\title[\civ\ absorbers in SDSS DR12]{Machine Learning Uncovers the
 Universe's Hidden Gems: A Comprehensive Catalogue of \civ\ Absorption Lines in SDSS DR12}

\author[R.~Monadi et. al.]{
Reza Monadi$^{1,2}$\thanks{E-mail: reza.monadi@email.ucr.edu (RM)},
Ming-Feng Ho$^{1}$, Kathy L.~Cooksey$^{3}$, Simeon Bird$^{1}$.
\\
$^{1}$University of California, Riverside, U.S.A.\\
$^{2}$California Polytechnic State University, San Luis Obispo, U.S.A.\\
$^{3}$University of Hawai`i at Hilo, Hilo, U.S.A.}

\date{Accepted 2023 September 18. Received 2023 September 17; in original form 2023 April 28}

\pubyear{2023}

\begin{document}
\label{firstpage}
\pagerange{\pageref{firstpage}--\pageref{lastpage}}
\maketitle

\begin{abstract}
We assemble the largest \civ\ absorption line catalogue to date, leveraging machine learning, specifically Gaussian processes, to 
remove the need for visual inspection for detecting \civ\ absorbers. The 
catalogue contains probabilities classifying the reliability of the absorption system within a quasar spectrum. 
Our training set was a sub-sample of DR7 spectra that had no detectable \civ\ absorption in a large visually 
inspected catalogue. We used Bayesian model selection to decide between our continuum model and our absorption-line models. 
Using a random hold-out sample of 1301 spectra from all of the 26,030 investigated spectra in DR7 \civ\ catalogue, we 
validated our pipeline and obtained an 87\% classification performance score. 
We found good purity and completeness values, both $\sim 80\%$, when a probability of $\sim 95\%$ is 
used as the threshold. Our pipeline obtained similar \civ\ redshifts and rest equivalent widths to our training set. 
Applying our algorithm to 185,425 selected quasar spectra from SDSS DR12, we produce a catalogue of 113,775 \civ\ doublets
 with at least 95\% confidence. Our catalogue provides maximum a posteriori values and credible intervals for \civ\ redshift, column density, and 
 Doppler velocity dispersion. We detect \civ\ absorption systems with a redshift range of 1.37--5.1, including 33 systems with a redshift larger 
 than 5 and 549 absorbers systems with a rest equivalent width greater than 2 Å at more than 95\% confidence. 
 Our catalogue can be used to investigate the physical properties of the circumgalactic and intergalactic media. 


\end{abstract}




\section{Introduction}
Metals, elements heavier than helium, are formed in the hearts of
massive stars and recycled into the interstellar medium (ISM) by
supernovae and stellar winds. Ultimately some of these metals are
transported into the circumgalactic medium (CGM) or even intergalactic medium (IGM).
Studying metals in the CGM sheds light on the complex, interconnected processes of accretion, feedback, and continual
recycling \citep{WhitePaper2020}. 
Measurements of the abundance of metals in the
Universe over time allow us to study the
cycling of baryons through galaxies and, thus, the
formation and evolution of galaxies \citep{cgmPaper, baryonCycle}.

Quasar absorption lines enable 
us to measure the abundance of elements and their ionization states
within the intergalactic gas.
Particularly useful is the \civ\ $\lambda\lambda 1548,1550$ 
doublet.
This doublet is caused by a strong transition of an abundant metal that
redshifts into optical bands at  $z \sim 1.5-5.2$, with lower redshifts
observable in the UV \citep[e.g.,][]{C10,shul2014, Hasan2} and higher in
the IR \citep[e.g.,][]{simcoeIR,weber2009,becker2009,davis2023}.
The rest wavelengths of the \civ\ doublet make it
detectable outside the \ion{H}{1} Ly$\alpha$ forest.
Moreover, \civ\ has an unsaturated doublet
ratio of $2:1$ for $W_{r,1548}:W_{r,1550}$, easing automated
line detection methods \citep{ChurchillBook}.

\civ\ is a resonance line doublet that is  useful for
studying many physical properties of the IGM and CGM over cosmic time.
It has been extensively studied;
here we will provide an abbreviated overview, and the
interested reader is referred to \cite[and references
therin]{baryonCycle} for a more comprehensive review.

%
Studying statistical properties  of \civ\ absorption systems, such as their
rest equivalent width distribution, sheds light  on all of the processes that contribute to the formation and propagation
of this metal throughout the IGM and CGM \citep{songaila2005, odorico2010, simcoe2011, Hasan1, Hasan2}.
The metallicity and enrichment history of the CGM was studied  using
the \civ/\ion{H}{I} line ratio \citep{barlow1998, Ellison2000}.
The ratio of \civ\ to other metal lines can
constrain the ionization state of the IGM \citep{boks2015}.
Also the ratios of different carbon ions (\ion{C}{II}\slash\civ) can be used to infer the
ionization state of the absorbing gas in the IGM at a redshift where neutral hydrogen absorption is saturated
\citep{cooper2019}.
One can measure or constrain the temperature and kinematics of \civ\ absorbers to analyze
the physics of the IGM \citep{rauch1996, Appleby2023}.
The study of the characteristics of metal lines, such as \civ, offers valuable information
for developing models of contamination in baryon acoustic oscillation measurements of the Lyman-$\alpha$ forest
\citep{Yang2022}. 
The auto-correlation (clustering) of \civ\ absorbers systems will
constrain the IGM metallicity and enrichment topology
\citep{sargent1988,petit1994, chen2000, scannapieco2006, Tie2022}.
Close quasar-galaxy pairs  connect \civ\ absorbers to galactic halos and
provide  a tool for studying galaxy evolution
\citep{Adel2005, bordoloi2014, rubin2015, burchett2015, burchett2016}.

\civ\ absorbers have been observed at $z > 5$, probing the tail end of
 the reionization epoch
\citep{becker2009, weber2009, simcoeIR, odorico2013,codoreanu2018, doughty2023}. 
Combining \civ\ ionisation data with other ions like \ion{Si}{IV} and \ion{C}{II} at high redshift, 
one can study the reionization-epoch ultra violet background's slope \citep{Finlator2016}.
Comparing \civ\ observational data with simulations shows some discrepancies in the production of carbon,
motivating more detailed observations and improved theoretical models \citep{Finlator2020}.

Most relevant to our current work, \cite{C13} detected strong \civ\ absorbers
in the low signal-to-noise spectra of the Sloan Digital Sky Survey (SDSS) \citep{sdssdr7, Eisen11}.
\footnote{\cite{Chen2014} assembled a \civ\ catalogue from SDSS DR9 quasar spectra. However, we did not use their candidate absorbers as a detailed comparison to previous \civ\ catalogues was missing.}
On the theory side, \civ\ has been associated with enriched gas surrounding
galactic halos in cosmological simulations \citep{Haehnelt96,c4FootPrint}.

The above surveys and catalogues of \civ\ were assembled by
visual inspection of quasar spectra by trained astronomers, sometimes
supplemented by template fitting to discover candidate absorbers.
However, this visual inspection is prohibitively time-consuming 
with the large size
of modern quasar surveys. The largest \civ\ catalogues are from SDSS:
\cite{C13} used Data Release (DR) 7 and \cite{Chen2014} used DR9.
The visually inspected quasar catalogue of SDSS DR12 contains 185,541 quasars \citep{sdssdr12ross},  which
can potentially have \civ\ absorbers.
The upcoming  Dark Energy Spectroscopic Instrument \citep[DESI][]{desi}
will obtain spectra for more than 30 million galaxies and quasars.
DESI will observe more than ten times the number of galaxies observed by SDSS and 
$\sim10^7$ quasars. 
Leveraging the increase in quasar spectra  for \civ\ studies is best served by an automated detection algorithm.
Neural networks are machine learning methods used for
detecting different absorption lines in quasar spectra, e.g.:
\ion{C}{II} \citep{CIIcnn},  Lyman-$\alpha$ forest \citep{Lyman-alphaCNN},
\civ\ broad absorption lines in SDSS \citep{BALcnn},
damped Lyman-$\alpha$ absorbers \citep{dlaCNN, DESIcnn} and \ion{Mg}{II}
\citep{MgIIdNN}. Our catalog is the first machine
learning based \civ\ quasar absorption line catalogue. 
Our use of Gaussian processes instead of neural
networks allow for reduced training times and easier estimation of uncertainty.

SDSS DR12 contains the largest extant quasar spectral catalogue with \emph{visually verified redshifts}. However, it has a relatively low spectroscopic resolution and a low median signal-to-noise ratio (SNR).
This makes the detection of an absorption line, like the \civ\ doublet, quite
challenging.
However, our Bayesian approach based on Gaussian processes is capable of extracting
reliable 
information even from noisy data.

Our automated \civ\ detection pipeline is based on the technique for detecting Damped Lyman-$\alpha$ absorbers (DLAs) 
from \cite{romanDLA}, which was extended to multiple absorbers by \cite{mfDLA}. A Gaussian process model with a bespoke learned kernel is built for the quasar spectrum in the absence of absorption, and Bayesian model selection is used to determine whether an absorber is preferred over the no-absorption (i.e., continuum) model given the quasar instrumental noise.
The pipeline is built using a Bayesian framework, allowing us to make probabilistic statements even about the noisiest observed data. Detection probabilities can be used to further refine the catalogue to increase purity or completeness.
Furthermore, as a fully Bayesian pipeline, it provides a
posterior distribution for the column density, redshift, and Doppler
velocity dispersion   for each absorber.

The rest of this paper is structured as follows. In Section \ref{sec:data}, we summarise the data 
we used for different stages in our pipeline. In Section \ref{sec:method}, we detail
the mathematical
framework for obtaining our absorption models, our Gaussian
process model for quasar emission, and our Bayesian approach to
search for absorbers in the quasar spectra. We validate our approach
by testing our algorithm in a hold-out sub-sample of our training set
in Section \ref{sec:validation}. The resulting \civ\ catalogue is presented
and discussed in Section~\ref{sec:results}. We summarise and discuss potential
future applications of our catalogue in Section \ref{sec:conclusion}.

\section{Data}
\label{sec:data}

Our primary dataset was SDSS quasar spectra; we followed \cite{C13} in designating
 quasars with their spectroscopic modified Julian date, fibre identification number, and plate number.
We trained our absorption-free model on a subset of SDSS DR7 \citep{sdssdr7} filtered to
avoid \civ\ absorbers as detected by the so called ``Precious Metals'' (PM) catalogue \citep{C13}.\footnote{We obtained
the list of spectra from \url{igmabsorbers.info} 
and downloaded the spectra from \href{http://das.sdss.org/spectro/1d_26}{http://das.sdss.org/spectro/1d\_26}}
The PM catalogue did not search for absorbers in spectra that did not meet certain criteria  (see Table 1 in \citeauthor{C13} 
for more details). Excluded were spectra of a broad absorption line quasar, 
spectra with insufficient wavelength coverage fo \civ, 
and
spectra with low median SNR ($\langle S/N\rangle<4\,{\rm pix}^{-1}$).
Our training set is based on the PM \civ\ catalogue, so starting from  SDSS DR7, we also
exclude quasar spectra not searched by \cite{C13}. The initial DR7 quasar
catalogue contains 105,783 quasars, of which 26,030 were
searched for \civ\ absorption. Our training set further
excluded the 10,861 spectra which contain one or more \civ\ absorbers
in the PM catalogue.
Our null model was thus  trained on 15,169 ``\civ-free'' spectra, meaning
spectra that either were not found as a \civ\ \textit{candidate} (as defined by \cite{C13})
or did not pass the \textit{visual verification} check.

Before training a continuum model on all of the 15,169 spectra in \cite{C13},  we train a
number of candidate continuum models on 95\% of our training set and then validate these
candidate continuum models on a
random hold-out sample of 5\% of all searched spectra in the DR7 catalogue, which contains 1301 quasars. This is
our \emph{validation set} that we used as a tool to find the optimum values of the parameters needed
to train a candidate continuum model. These tuning parameters include: flux normalization wavelength range,
the minimum number of non-NaN pixels in a training spectrum,
the dimension of the covariance matrix (see Equation \ref{eq:K}), etc. After applying our pipeline on the
validation set, we assessed the performance of the classification (i.e. classifying a
given spectrum as having \civ\ absorber(s) or otherwise) using the PM catalogue as a ``ground truth''.
We found the best candidate continuum model by maximising the classification score
(see Section \ref{sec:roc}) and purity/completeness
 (see Section \ref{sec:purity_completeness}).
 At this point, we took the parameters of the best candidate continuum model and
built our final model from all of the 15,169 ``\civ-free'' DR7 spectra investigated in the PM catalogue pipeline.

We applied our algorithm on a subset of the SDSS DR12 quasar catalogue \citep{sdssdr12alam}
to build our new \civ\ catalogue. We chose our working quasar sample starting from the
SDSS-DR12 quasar catalogue.\footnote{\url{http://data.sdss3.org/sas/dr12/boss/qso/DR12Q/DR12Q.fits} }
We kept only quasars with rest-frame wavelength coverage between 1310~\AA--1548~\AA, 
the region of potential \civ\ absorption (avoiding both the Ly$\alpha$ forest and the potential for false positives of \civ~from \ion{O}{I} $\lambda1302$ or \ion{Si}{II} $\lambda1304$). 
This means quasars with redshifts satisfying 1310~\AA(1+$\zqso$)> 3650~\AA\ (or $\zqso>1.7$)
and 1548~\AA(1+$\zqso$) < 10400~\AA\ (or $\zqso<5.7$).
We removed detected broad absorption line quasars (BAL) using
the SDSS BAL
catalogue\footnote{\url{http://data.sdss3.org/sas/dr12/boss/qso/DR12Q/DR12Q_BAL.fits} }.
After these selections, we downloaded the list of
quasar spectra
from the SDSS-III Baryon Oscillation
Spectroscopic Survey Science Archive
Server\footnote{\url{https://data.sdss.org/sas/dr12/boss/spectro/redux/} }.

We converted all observed spectra to
the emission rest-frame using the visually inspected quasar
redshift estimate from the SDSS
pipeline, which we assume to be exact.\footnote{We used \texttt{Z\_VI}, column 8 of SDSS DR12
quasar catalogue.}
Missing or otherwise masked flux values (e.g., from a bad pixel) were denoted by \texttt{NaN} and were not used in our pipeline.


\section{Method}
\label{sec:method}

We  modified the pipeline introduced in \cite{romanDLA} and \cite{mfDLA} to look for \civ\ absorbers in SDSS DR12. We learnt an \textit{a priori}
distribution for the shape of the quasar emission spectra without \civ\ using SDSS DR7 spectra classified by the PM catalogue from \cite{C13}.
The null model, $\model_{N}$,  
was learned from SDSS DR7 spectra identified as `non-detection' (i.e., no \civ\ \textit{candidate} in the PM study). 
Each iteration, we did a Bayesian model selection between the null model, a model for a \civ\ doublet model ($\model_D$), and a model for an `interloper' singlet absorption line ($\model_S$) 
to compute the posterior probability of \civ\ absorption.
We searched for up to seven 
\civ\ absorbers in each spectrum, reporting probabilities for each.
We stopped at seven absorbers as only 4 spectra among 26030 investigated spectra in the DR7 \civ\ catalog
contained seven absorbers and none contained more.

\begin{figure}
\centering
\includegraphics[width=\linewidth]{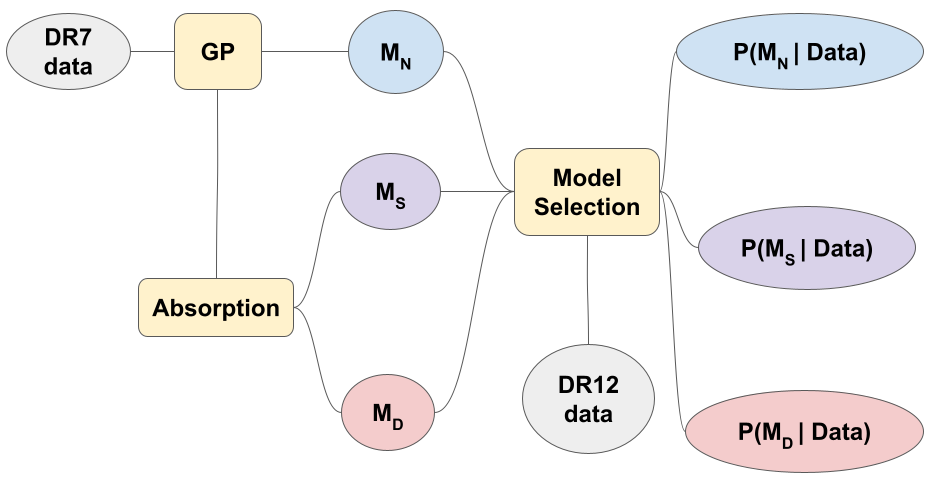}
\caption{This is a flow chart for our pipeline.
 Training spectra from SDSS DR7 are used to train a Gaussian process
 kernel with which to model the quasar continuum (i.e., null model, $\model_{N}$).
Analytic Voigt profiles are used to construct models for absorption from a \civ\ doublet ($\model_D$) or a generic singlet absorber ($\model_S$).
Conditioning on DR12 spectra produces a posterior probability estimate for each model that
can be used to decide if there is a \civ\ absorber in the given spectrum or not.
Moreover, for the absorber models, $\model_D$ and $\model_S$,   we have a posterior distribution
for each model parameter: absorber redshift, Doppler velocity dispersion for the absorption profile,
and the absorber column density.}
\label{fig:flowChart}

\end{figure}

There were six main changes since \cite{mfDLA}. 
First, the absorption profile was updated to model a \civ\ doublet, instead of a DLA. Second, a model for singlet line absorbers was introduced, which serves a similar role to the sub-DLA model in \cite{mfDLA}. Without this singlet absorption line model, the pipeline produces excessive false positives, as it has no other way to match absorption except a \civ\ doublet.
Third, in addition to sampling absorber redshift and column density, we sampled the Doppler velocity dispersion, which allows more accurate fits.
Forth, we no longer model the Lyman-$\alpha$ forest in the null model, as it does not overlap our \civ\ absorption region.
Fifth, instead of a fixed instrumental broadening profile, we used the
reported Gaussian wavelength-dependent dispersion from SDSS.\footnote{Column 6 of the fits files of SDSS spectra, see the SDSS
\href{https://data.sdss.org/datamodel/files/BOSS_SPECTRO_REDUX/RUN2D/spectra/PLATE4/spec.html}{data model}.}
Sixth, we report an individualised probability for each absorber we detect, rather than the
joint probability of observing at least a certain number of absorbers, as in \cite{mfDLA}.
Figure \ref{fig:flowChart} shows an overview of our pipeline, as described in this Section.

Section~\ref{sec:data} described our
initial training data, a subset of SDSS DR7. Section~\ref{sec:absorption}
explains the Voigt-profile model 
for any absorber detection.  
Section~\ref{sec:null_model}
summarises our null (aka absorption-free or continuum) model,  $\model_N$, for the quasar emission function, which
uses a bespoke Gaussian Process kernel. 
Section~\ref{sec:civ-model}
describe two analytic absorption models, $\model_S$ and $\model_D$, which 
are generated by convolving $\model_N$ with a singlet or
doublet Voigt profile, respectively. In addition, we need model
priors, $Pr(\model)$, for each model (see
Section~\ref{sec:modelpriors}). The model likelihood is
discussed in Section~\ref{sec:model_likelihood}.
Section~\ref{sec:multiciv} explains our technique for deciding how many \civ\ absorbers to search for.


\subsection{Absorption function}
\label{sec:absorption}

Voigt profiles are useful for modelling the absorption effect in the
observed spectrum of an emitting source such as quasars \citep{ChurchillBook}. A
Voigt profile is given by:
\begin{equation}
\begin{split}
\phi(\rm{ v}, \sciv, \gamma_{\ell u})
&=\\
\int \frac{d {\rm v}}{\sqrt{2\pi}\sciv} &\exp{(-{\rm v}^2 / 2\sciv^2)}
\frac{4 \gamma_{\ell u}}{16 \pi^2 [ \nu - (1 - {\rm v}/c)\nu_{\ell u} ]^2 + \gamma_{\ell u}^2},
\end{split}
\label{eq:voigt}
\end{equation}
which is a convolution between Lorentzian and Gaussian profiles.
The former computes the natural broadening and the latter thermal
broadening \citep{DrainBook}. The velocity, $\rm v$, in Equation \ref{eq:voigt} is
given by:

\begin{equation}
\rm v(\lambda) = c\bigg(\frac{\lambda}{\lambda_{\ell u}(1+\zciv)} -1\bigg).
\end{equation}
A negative (positive) velocity refers to a position in $\lambda$-space that is red-ward (blue-ward) of the
observed \civ\ absorption in $\lambda_{\ell u}(1+\zciv)$. The Lorentzian broadening contribution is:
\begin{equation}
\gamma_{\ell u} = \frac{\Gamma \lambda_{\ell u}}{4\pi},
\end{equation}
where $\Gamma$ is the damping constant.
The Doppler velocity dispersion for a \civ\ absorber, $\sciv$, is:
\begin{equation}
\sciv = \sqrt{\frac{kT}{6m_p+6m_n}},
\label{eq:sigma}
\end{equation}
where $k$, $T$, $m_p$ and $m_n$  are the Boltzmann constant, gas temperature,
proton mass, and neutron mass, respectively. The Doppler velocity dispersion  controls the
width of the absorption profile as a function of temperature.
For the \civ\ doublet at
$\lambda=1548$~\AA, $\Gamma =  2.643\times 10^8$ s$^{-1}$ and
for $\lambda=1550$~\AA,\ $\Gamma = 2.628\times 10^8$ s$^{-1}$.
Lorentzian broadening is thus small
($\gamma_{\ell u}/\sciv\sim 0.01$ for $T\sim10^4$K) and
the Voigt profile is close to Gaussian.

The optical depth, $\tau$,
itself is a function of
observed frequency ($\nu = c/\lambda$) given: 
absorber column
density $N_{\civ}$ which controls the depth of the profile,
absorber redshift $\zciv$ which sets the wavelength where
we observe the absorption, and  Doppler velocity dispersion  $\sciv$.
 The optical depth is given by:
\begin{equation}
\tau_{\ell u}(\lambda; \zciv, \nciv, \sciv)
=  \frac{\nciv\pi e^2 f_{\ell u} \lambda_{\ell u}}{m_e c}
\phi(\rm v(\lambda), \sciv, \gamma),
\label{eq:optDepth}
\end{equation} 
where $c$ is the speed of light, $e$ is the elementary charge, $m_e$ is the mass of the electron and
$\lambda_{\ell u}$ is the transition wavelength for the lower state ($\ell$) and the upper state ($u$) 
and $f_{\ell u}$ is the oscillator strength of the transition.
Using spectroscopic notation \citep{specBook}, the 1548~\AA\ absorption
line is a transition from
$2^2 {S}_{\frac{1}{2}}$ to $2^2 {P}_{\frac{1}{2}}^{o}$  and the $1550$~\AA\ absorption line is
a transition from $2^2 {S}_{\frac{1}{2}}$ to $2^2 {P}_{\frac{3}{2}}^{o}$.
The absorption profile  is related to the optical depth via:
\begin{equation}
\label{eq:a}
a_{\ell u}(\lambda; \zciv, \nciv, \sciv) = \exp{(-\tau_{\ell u}(\lambda; \zciv, \nciv, \sciv))}\,,
\end{equation}
where the $\ell u$ subscript can refer to either $1548$\,\AA~or $1550$\,\AA\, transitions.
The doublet model $\model_D$ will be built by convolving the null model with an absorption profile that considers
both $1548$\,\AA~or $1550$\,\AA. The singlet model $\model_S$, on the other hand, only considers the $1548$\,\AA\ transition.

SDSS resolution is insufficient for detailed modelling of \civ\ absorption
systems as is done with high-resolution spectra \cite[e.g.~][]{Hasan1}.
Indeed, strong \civ\ absorption at SDSS resolution can be
reasonably modelled by a single Voigt profile with appropriate choice
of  $\zciv$, $\nciv$, and $\sciv$,
as we do in this work
(see Section \ref{sec:model_likelihood}).
We acknowledge that the same absorption at
higher resolution would reveal finer structure
and require multiple Voigt profiles, with
different combinations of $\zciv$, $\nciv$, and $\sciv$\ that
would be strong constraints on the physical conditions of the gas
giving rise to the absorption. The $\nciv$\ and $\sciv$\ values returned
by our algorithm may not be as tightly constrained as the $\zciv$ measurements.
Remember that $\nciv$\ and $\sciv$\ control the Voigt profile shape
in our absorption model that is compared to the observed flux deficit in the SDSS spectra (see Table \ref{tab:gpTab}).

\subsection{Quasar emission function}
\label{sec:null_model}

The physics of quasar emission is not fully understood, and
there is considerable variety in observed quasar spectra.
Thus we used an empirical model for the quasar emission function (aka continuum) 
based on the observed spectra. We  modelled the emission function of a quasar,  $f$, in the absence of any
absorption (including
\civ\ absorption) using \emph{Gaussian processes} that  generate a
distribution over functions. Gaussian Processes 
result from a
generalisation of a multivariate Gaussian distribution to
infinite domains \citep{GPbook}.
As the standard library of kernels is insufficiently flexible
to model the complicated correlations between different emission lines
in a quasar spectrum, we used a
customised  kernel learned directly from the training set.\footnote{
Our training set consisted of all of the spectra investigated in the PM
\civ\ catalogue and classified as not containing \civ\ absorbers.}
We described the training set in Section \ref{sec:data}.

\begin{figure}
\includegraphics[width=\linewidth]{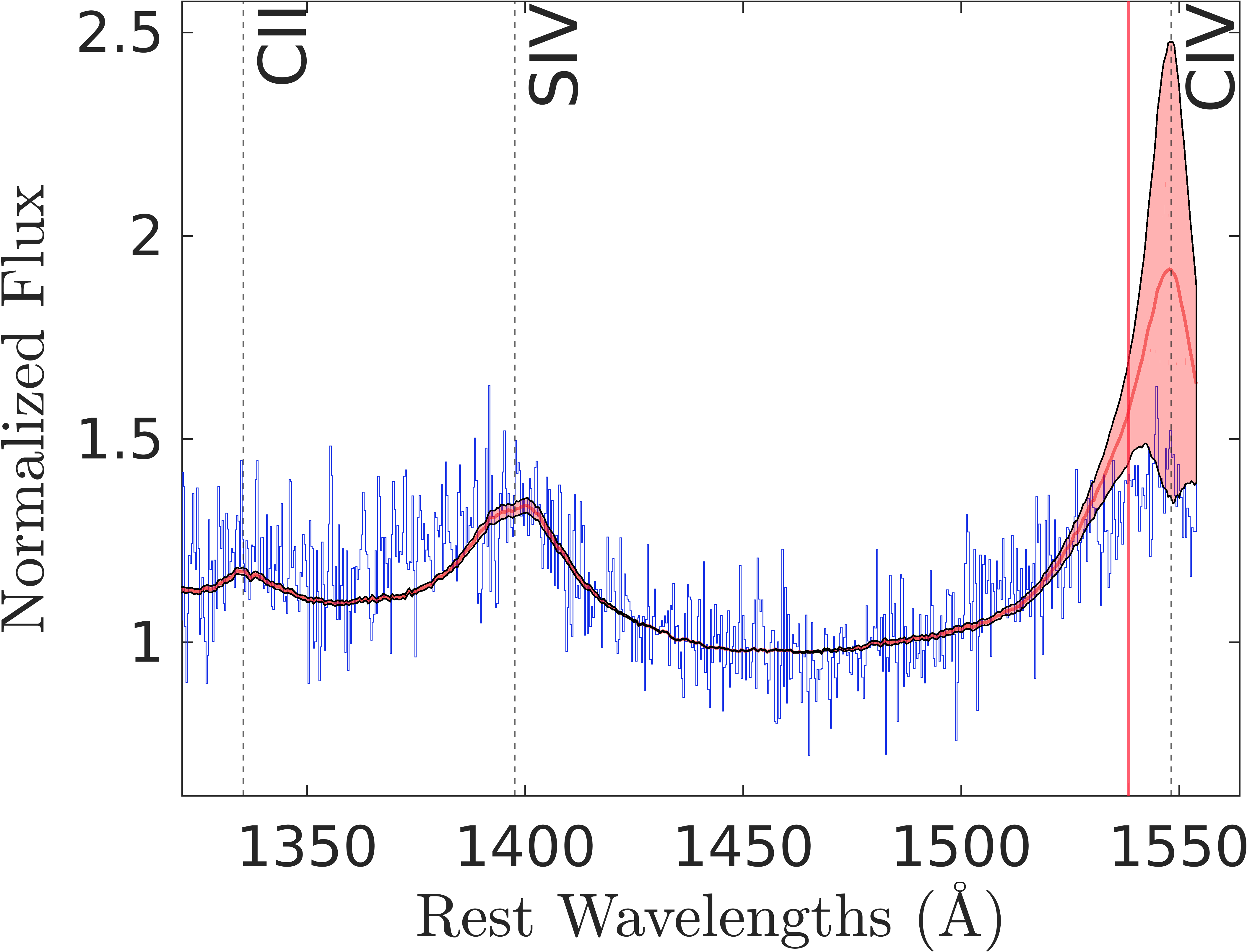}

\caption{An example learned quasar emission function (red curve) with the normalised observed smoothed flux (blue curve).
The shaded red region shows 1$\sigma$ uncertainties.
The SDSS DR7 quasar has QSO-ID: 51630-0266-280 and redshift $2.57$.
    Note that we search for absorbers starting 3000\,\kms\ red-ward of the quasar's redshift (shown by the solid red vertical line), so the moderate failure to match the quasar \civ\ emission line in this case does not lead to an artificial preference for \civ\ absorption. Prominent emission lines are marked by dashed vertical lines.
}
\label{fig:mu}
\end{figure}

Here we briefly
summarise the technique. Our model was similar to \cite{romanDLA}
where  the process of
obtaining a Gaussian process model for quasar emission spectra is described in more detail.
However, unlike \cite{romanDLA}, we did not model the Lyman-$\alpha$ forest
as we were looking for \civ\  absorbers outside of the forest.
We trained a \civ-free model between 1310~\AA\ and
1555~\AA, which produced the best results during the validation phase.
This range is close to the rest frame  \civ\  absorption wavelength
searched in the PM catalogue.
Figure \ref{fig:mu} shows an example learned quasar continuum together
with the observed flux and noise.

Even from low SNR spectra, our method extracts some statistical information, so we do not enforce a minimum SNR in our search. 
Our pipeline naturally gives low likelihoods to low-SNR spectra  during the training.  
We can completely specify a Gaussian process 
by its mean and correlation functions (analogous to the first two moments of a Gaussian distribution).
We specify the mean function $\mu$ by:
\begin{equation}
\mu(\lambda) = \langle y(\mathbf{\lambda})  \rangle, \label{eq:mu}
\end{equation}
where $\lambda$ is the rest-frame wavelength and $y(\lambda)$ is
the observed rest-frame flux for the training-set spectra, after
applying a mask for missing pixels; angle brackets ($\langle \rangle$)
denote an average over wavelengths.
Before computing this average over the training set, we have normalised the quasar flux
and the flux variance  so that they
have a median value of unity in the normalisation range.
This normalisation was needed
so that the Gaussian process model is insensitive to variations in (observed) quasar brightness.
We chose the range from 1420~\AA \ to 1475~\AA \ as it contains no prominent emission lines \citep{Zhu13, hamann17, Monadi21}.
We also confirmed empirically that this normalization range produces the best score when applied to our validation set.
We remind the reader that the validation set is a random subset (1301 spectra) of all candidate DR7 spectra in
the PM catalogue (see Section \ref{sec:data}). 


The Gaussian process covariance function describes the correlation between
flux values at two separate wavelengths, $\lambda$ and $\lambda^{\prime}$. 
Most applications of Gaussian processes assume a simple kernel for the covariance, such as the exponential squared kernel \citep{GPbook}.
However, the complex correlation between features in
quasar continua is hard to describe using the simple/standard
covariance functions like the radial basis function. 
Instead, our algorithm 
directly learned a covariance function:
\begin{equation}
K(\lambda, \lambda') =  {\rm cov} [f(\lambda), f(\lambda')],
\label{eq:covFunction}
\end{equation}
by considering all of the cumulative
information
contained in the spectra of our training set:  all of the flux measurements and noise measurements given at the observed wavelengths.\footnote{The third column of the SDSS fits tables
for observed spectra contains inverse noise variance $(\sigma(\lambda))^{-2}$.}  We need to  maximise the joint likelihood of generating the whole training set
given that the underlining model is the null model (i.e. absorption-free).
We assume our observations (i.e. flux and noise given at each observed wavelength in the training set) are independent and
drawn from a Gaussian distribution with width corresponding to the observed noise of the SDSS pipeline.
Next we maximise the likelihood (see section 5.3 of \citep{romanDLA} for
details) and learn the quasar mean function (Equation \ref{eq:mu}) and quasar covariance function (Equation \ref{eq:covFunction}).
Optimising this joint likelihood function was done using \texttt{minFunc}: a Matlab function for unconstrained optimization of differentiable real-valued multivariate functions using line-search methods.\footnote{\href{https://www.cs.ubc.ca/~schmidtm/Software/minFunc.html}{https://www.cs.ubc.ca/~schmidtm/Software/minFunc.htm}}

\begin{figure}
\includegraphics[width=\linewidth]{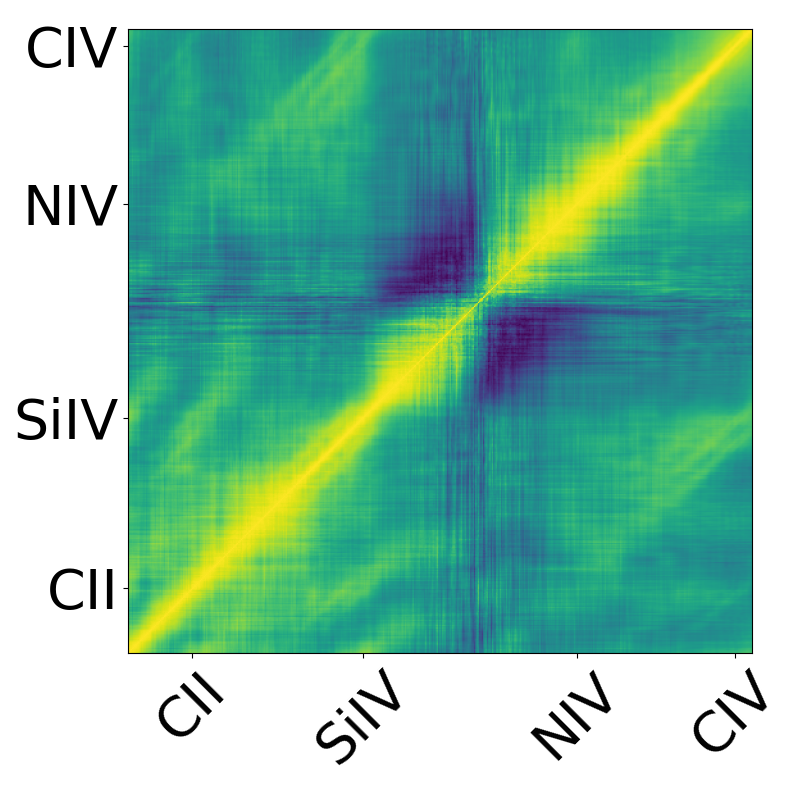}
\caption{Learned covariance matrix $\mathbf{K}$ (see Equation \ref{eq:covFunction} and Equation \ref{eq:K})
for our null (continuum) model. This matrix is built up by considering
the observed flux and noise from our \civ-free training set (see Section \ref{sec:data}). Brighter pixels show stronger correlations and darker regions weaker ones. The wavelengths of prominent emission lines are labelled. The bright diagonal implies stronger correlations between pixels at smaller wavelength separation.
}
\label{fig:K}
\end{figure}

We binned quasar spectra linearly in wavelength, from $1310\!-\!1555$~\AA, 
with a bin size of $\Delta \lambda$. This  gave us the number of bins as:
\begin{equation}
N_{\rm bin} = \frac{1555 - 1310}{\Delta \lambda}.
\end{equation}
If we input the binned wavelength grid, $\boldsymbol{\lambda}$, to Equation \ref{eq:mu}
we get the learned mean vector $\boldsymbol{\mu}$, with
$N_{\rm bin}$ elements. The covariance matrix, $\mathbf{K}$, an $N_{\rm bin}\times N_{\rm bin}$ matrix, is calculated on two discretized wavelength grids, $\boldsymbol{\lambda}$ and
$\boldsymbol{\lambda'}$, using Equation \ref{eq:covFunction}.
A very fine $\Delta \lambda$ is not desirable because it increases the size
of $\boldsymbol{\mu}$
and $\mathbf{K}$ 
and thus is more computationally
expensive.
On the other hand a coarse $\Delta\lambda$ cannot capture
enough information from the quasar
spectra. The optimum $\Delta\lambda$ in \cite{romanDLA} and \cite{mfDLA}
was 0.25~\AA. We empirically found that $\Delta\lambda=0.5$~\AA\
is the optimum value for the redder spectral region we examine here which gives us
$N_{\rm bin}=490$.
Without further structural assumptions on $\mathbf{K}$, our algorithm would have to
learn a matrix of $N_{\rm bin}^2 \sim 2.4 \times 10^5$ elements.
To circumvent this, we used a low rank decomposition:
\begin{equation}
\mathbf{K} = \mathbf{M}\mathbf{M}^{\top},
\label{eq:K}
\end{equation}
where $\textbf{M}$ is a $N_{\rm bin} \times k$ matrix, for any positive integer $k$.
Larger-$k$ models allow for higher fidelity modelling of $\mathbf{K}$.
Following \cite{romanDLA}, we set $k=20$. We also checked
$k=19$, $21$, and $22$, finding that our results were insensitive to this choice.
Figure \ref{fig:K} shows the learned covariance matrix.
This covariance matrix describes how likely the quasar emission spectrum is to vary around the mean spectrum. It encodes the information contained in the spectra of our training set, the ``\civ-free'' spectra from the PM catalogue.


Having learned the mean quasar vector $\boldsymbol{\mu}= \mu(\boldsymbol{\lambda})$ (see Equation \ref{eq:mu}) and 
the lower rank decomposition matrix $\mathbf{M}$ in Equation \ref{eq:K} which gives us the covariance matrix
$\mathbf{K}$ (Equation \ref{eq:covFunction}), we can write
the Gaussian processes model for the quasar
emission function, trained on the observed spectra, as a multivariate Gaussian distribution:
\begin{equation} 
p(f(\boldsymbol{\lambda}))  =  {\mathcal{GP}}(\mu(\boldsymbol{\lambda}), K(\boldsymbol{\lambda, \lambda'}))
=  \mathcal{N}(f(\boldsymbol{\lambda}); \mu(\boldsymbol{\lambda}), K(\boldsymbol{\lambda, \lambda'})),
\label{eq:conti}
\end{equation} 
where $\mathcal{GP}$ denotes a Gaussian process.
We remind the reader that a Gaussian process is a Gaussian distribution  over functions. 
Therefore, we can write the Gaussian process for the quasar emission function $f$ given our learned mean vector $\boldsymbol{\mu}$
and covariance matrix $\mathbf{K}$  as:
\begin{equation}
\mathcal{N}(f; \boldsymbol{\mu}, \mathbf{K}) = \frac{1}{\sqrt{(2\pi)^d \textrm{det}(\mathbf{K})}}\exp\Bigl( -\frac{1}{2} (f-\boldsymbol{\mu})^{\top} \mathbf{K}^{-1} (f-\boldsymbol{\mu})\Bigr),
\end{equation}
where $d$ is the dimension of the quasar emission function $f$.
\subsection{Absorption line models}
\label{sec:civ-model}
We want to find the probability of
a \civ\ doublet in the observed spectrum of a quasar given the observed rest-frame flux $\boldsymbol{y}( \boldsymbol{\lambda})$, under our null (aka absorption-free or continuum) GP model $\model_N$.
Our data  were composed of the observed wavelengths  $\boldsymbol{\lambda}$,
their corresponding
observed quasar flux $\boldsymbol{y(\lambda)}$, and their corresponding
observed noise $ \boldsymbol{\sigma(\lambda)}$.
We define the data as:
\begin{equation}
  \label{eq:data}
  \Data = \{\boldsymbol{\lambda; y( \lambda), \sigma(\lambda)}\}.
\end{equation}
Bayes' rule gives the \emph{model posterior}, the probability of each model given the data:
\begin{equation}
P(\model_i \vert \Data) =
\frac{P(\Data \vert \model_i)Pr(\model_i)}{
  \sum_j P(\Data \vert \model_j)Pr(\model_j),
}.
\label{eq:model_selection}
\end{equation}
We defined three models:
\begin{itemize}
\item $\model_{N}$ models the quasar continuum without
absorption (Equation \ref{eq:conti}).
\item $\model_{D}$ is a model containing
exactly one \civ\ doublet. $\model_D$ is built
by convolving $\model_N$ with the absorption
function (Equation \ref{eq:a}) for all observed wavelengths.
\begin{equation}
\model_{D} \rightarrow \textrm{convolve}(a_{1548,1550}(\boldsymbol{\lambda}), \model_N)
\label{eq:model_D}
\end{equation}
\item $\model_{S}$ is a singlet model containing exactly one
generic singlet absorption line.
For simplicity, we implemented $\model_S$  using the
same Voigt profile as $\model_D$ but including only the
1548\AA\ absorption line.
\begin{equation}
\model_{S} \rightarrow \textrm{convolve}(a_{1548}(\boldsymbol{\lambda}),  \model_N)
\label{eq:model_s}
\end{equation}
\end{itemize}
We added this singlet model, in addition to the \civ-free and \civ-doublet models, so that
our Bayesian framework is not forced to give a high probability of a \civ\ doublet
if there is a strong singlet line in the spectrum and nearby noise
happens to be similar to a \civ\ doublet.
For example, a broad singlet line like \ion{Si}{II} 1526,
\ion{Fe}{II} $\lambda1608$, or \ion{Al}{II} $\lambda1670$, 
can be mis-identified as a \civ\ doublet, if we have only
two models (i.e. $\model_N$ and $\model_D$).
The singlet model, $\model_S$, provides an alternative to both $\model_N$ and $\model_D$ for such lines.

\begin{figure}
\centering
\includegraphics[width=\linewidth]{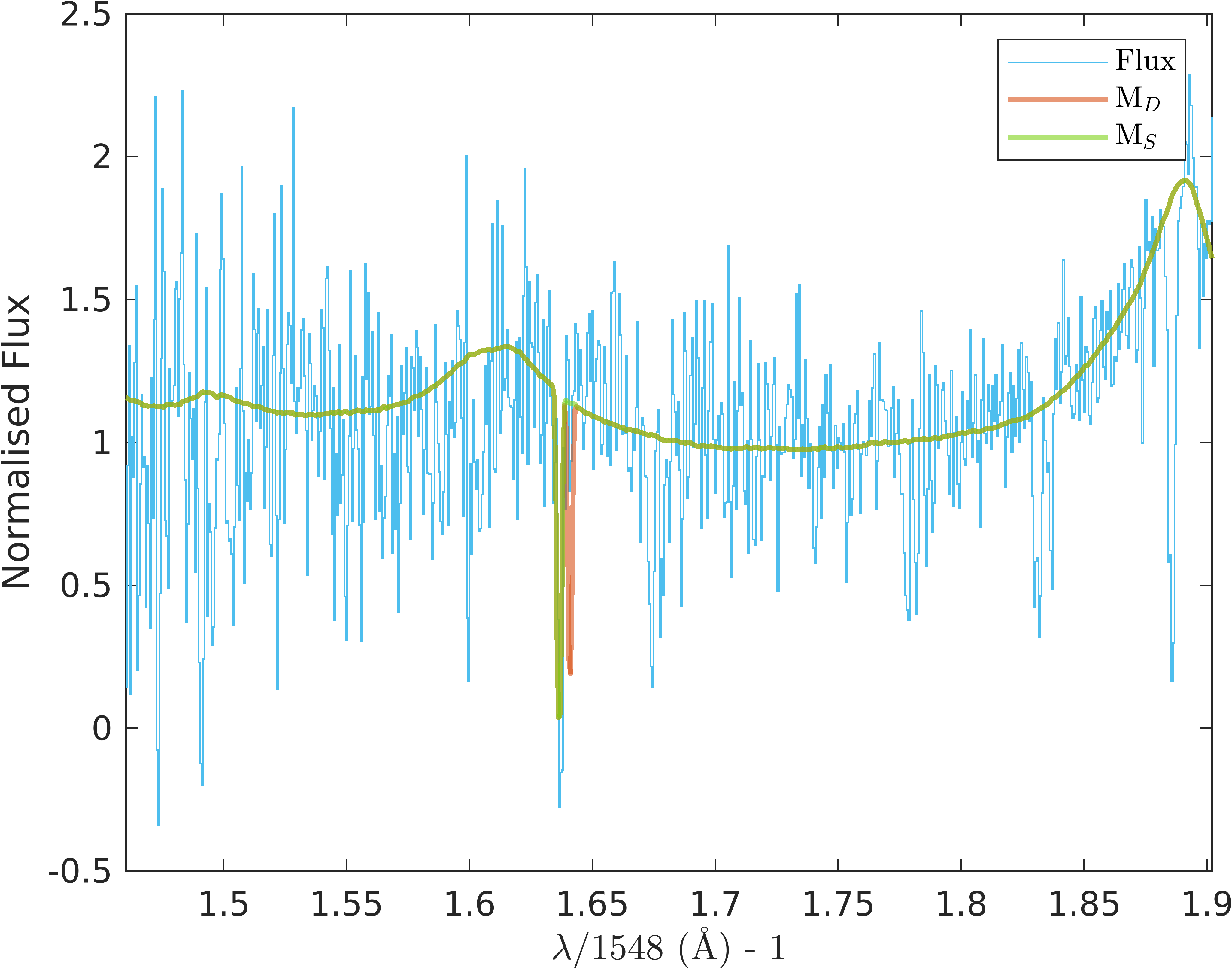}
\caption{The figure shows the spectrum of QSO-ID: 51608-0267-264 with $\zqso$=1.89 (blue) where the singlet model (green) is preferred over the \civ\ doublet model (red), which is in turn preferred over the null model.
If we did not have $\model_S$, our pipeline
would have incorrectly detected a \civ\ absorber at $\zciv = 1.635$.
}
  \label{fig:sngl}
\end{figure}

Figure \ref{fig:sngl} shows an example, the application of our pipeline to QSO-ID: 51608-0267-264 with $\zqso=1.89$. Here a noise fluctuation and a strong line happen to have a velocity separation similar to a \civ\ doublet.
For this spectrum, we have:
\begin{eqnarray}
\log(P(\model_{N} \vert \Data)) & = & -297.8216, \nonumber \\
\log(P(\model_S \vert \Data)) & = &  -250.1609, \mbox{and} \nonumber \\
\log(P(\model_D \vert \Data)) & = & -257.1906. \nonumber
\end{eqnarray}
Although the doublet model is not a very good fit, the
null model is even worse. Thus without the singlet
model, $\model_S$, our pipeline would incorrectly prefer
the doublet model and detect a \civ\ absorber. 
Note that these log-likelihoods are not normalised. We fit to the whole quasar spectrum so the number
 of data points (degrees of freedom) is large compared to the local change in the 
 spectrum around one absorber system. In addition, this particular spectrum has a second 
 absorber visible at $z=1.82$, which reduces the likelihood during the first absorber search. What is 
 important here is the difference between these log-likelihoods. In the process of Bayesian model selection
 we need the Bayes's factor\footnote{Bayes factor is the ratio between  posterior probability of two models. } which is proportional to the 
 difference between these log-likelihoods (See Equation \ref{eq:model_selection}). 

A sampling problem arises due to the low resolution of the SDSS spectrograph.
Real spectrographs measure the total integrated
flux across the spectral pixel. A simple estimate for this is to evaluate the Voigt profile
at the center of the pixel. However, at the low resolution of the SDSS spectrograph, this can be
a poor estimate, leading to unphysical doublet ratios. For this reason we compute the integrated flux by first evaluating the Voigt profile on a grid of pixels which is finer than the grid in the SDSS spectrum by a factor of $n_{\rm ave}$.
We found by experiment that the model accuracy does not improve for $n_{\rm ave} > 20$ sub-samples.

\subsection{Model priors}
\label{sec:modelpriors}
To calculate the model posterior (Equation \ref{eq:model_selection}),
we need model priors, $Pr(\model)$, for each of the three models.
We set priors for the \civ\ doublet, $\model_D$, using population statistics from our training set, the PM catalogue of \cite{C13}.
We counted the fraction of spectra with absorbers at $\zciv<\zqso - 30000 /c $,
where $c$ is the speed of light in \kms.
This small decrease in our upper 
limit for the
absorption redshift accounts for any possible error in estimating
the redshift from the SDSS pipeline.
For simplicity, we used the same prior for the singlet and doublet line models, i.e., $Pr(\model_S)=Pr(\model_D)$.
There are no single-line catalogues for these data, and using equal
priors ensures that whichever model is the best-fit will be used.

The prior for the \civ-free model can be obtained by:
\begin{equation}
Pr(\model_N(k~\civ)) = 1- Pr(\model_D(k~\civ)),
\label{eq:priorNull}
\end{equation}
where ``$k~\civ$'' denotes some integer number $k$ of \civ~systems.
We did not include $Pr(\model_S)$ in Equation \ref{eq:priorNull} to
enable a pointwise model comparison between $\model_D$, $\model_S$,
and $\model_N$. Especially when searching for multiple absorbers, our
main purpose is deciding the probability of detection or non-detection
of \civ\ absorbers in a spectrum.
Furthermore, the small shift in the normalization of model
priors is several orders of magnitude smaller than the effect
of normalising the model posteriors in Equation \ref{eq:model_selection}.

When searching for additional absorbers in spectra where there is already a detection, we use the prior probability of spectra with $(k-1)$~\civ\
absorbers having $k$ absorbers:
\begin{eqnarray}
  Pr(k~ \civ) &=& P(k~\civ\vert (k-1)~\civ) \nonumber \\
  &= &\frac{P((k-1) ~\civ \cap  k ~\civ)}{P((k-1) ~\civ)} \nonumber \\
  &= & \frac{P(k~\civ)}{P((k-1)~\civ)}\,.
\end{eqnarray}
The equality follows as the intersection between the set with
$k$ \civ\ and the set with  $(k-1)$~\civ\ will be the set of quasars with
$k$ \civ\ absorbers. $Pr(k~ \civ)$ is guaranteed  to be less than $1$, because
there are always fewer spectra with more absorption systems.

\begin{figure}
\includegraphics[width = \linewidth]{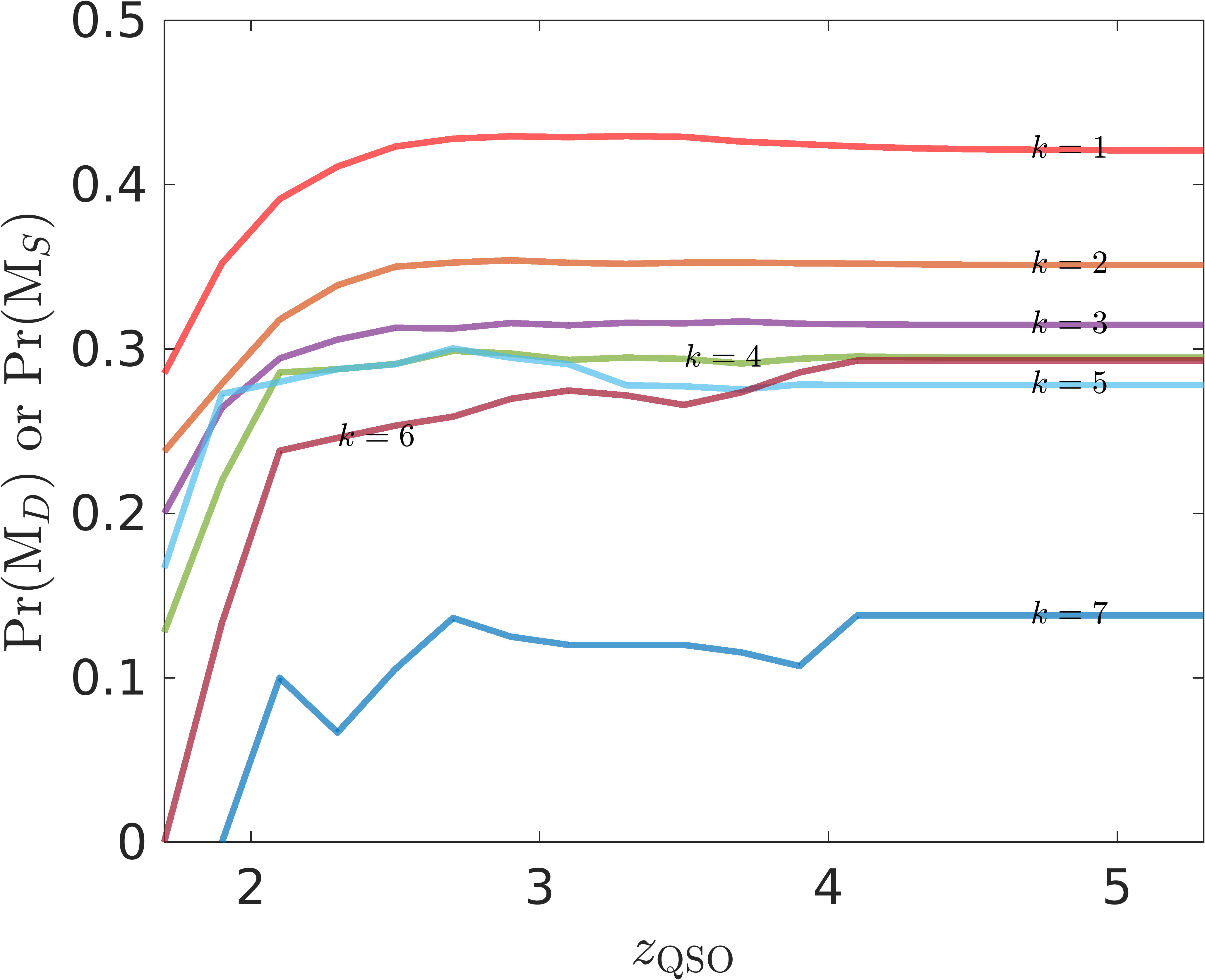}
\caption{Prior probability
for a spectrum containing $k$ \civ\ absorbers as a function of quasar redshift, for $k = 1\!-\!7$.
We use the average number of absorbers in the PM spectrum in our wavelength search range.
\civ~is \textit{a priori} more likely as $\zqso$
increases but reaches a plateau at $\zqso \sim 2.5\!-\!3$.
This is because the \civ\ wavelength coverage
is shorter for low $\zqso$ as the 1548~\AA\ emission
line pushes to the blue-end of the SDSS spectral range. Note that we assume the same prior for the singlet model for $k=1\!-\!7$.}
\label{fig:prior}
\end{figure}

Figure \ref{fig:prior} shows the $\model_D$
priors we used for different searches as a function of $\zqso$.
When the redshift increases, all of the priors reach a plateau  after $\zqso \sim 3$.
There is a  decrease from $Pr(1 ~ \civ)$ to the
subsequent priors so that $Pr(7 ~\civ) <15\%$.
\civ\ absorbers cluster \citep[eg.][]{Boksenberg2003}, so the prior for detecting $k$ absorbers in a spectrum given a redshift is larger than $Pr(1~\civ)^k$: when there is
no clustering and the absorbers are perfectly independent.

\subsection{Model likelihood}
\label{sec:model_likelihood}

The model likelihood, $P(\Data \vert \model)$, in Equation \ref{eq:model_selection} is
the probability that the observed data, $\Data$, have
been generated by a considered model, $\model$, after marginalising
the model parameters.
We do the marginalisation over a prior distribution for each parameter in the model:
\begin{equation}
P(\Data \vert \model) =
\int P(\Data \vert \model, \theta) P(\theta \vert \model) d\theta.
\label{eq:marginal_params}
\end{equation}
Here $P(\Data \vert \model, \theta)$ is the likelihood of the spectra
being generated
by model $\model$, if the model has a certain set of parameters $\theta$. We
use the prior probability distribution of $P(\theta \vert \model)$ from Equation \ref{eq:marginal_params}
to integrate out all of the possible $\theta$ and obtain a
parameter-independent model likelihood. The null model $\model_{N}$ has no
free parameters, but $\model_D$ and $\model_S$ have three free parameters each:
1) absorption redshift ($\zciv$), 2) column density of $\civ$ ($\nciv$), and 3)
Doppler velocity dispersion  ($\sciv$). As mentioned in Section \ref{sec:absorption}, it is
sufficient for our purposes to model an absorption line at SDSS resolution with
a single Voigt profile (defined by $\zciv$, $\nciv$, and $\sciv$) and
use these values to measure a rest equivalent width; however, only redshift
are well-constrained by the data.

Our algorithm fits a 
single Voigt profile, instead of a combination of Voigt profiles with low velocity separations, even when we have a blended/complex absorption system.
To obtain a reasonable fit with a single Voigt profile, we include large Doppler broadening velocities, which do not necessarily reflect the
temperature of the absorbing gas. This keeps our algorithm simpler by sacrificing the reliability  of Doppler 
velocity dispersion measurements. In a follow up work we will modify the algorithm so that it can measure the 
temperature as well. 


We need to have priors for each of these
parameters to
perform the integral in Equation \ref{eq:marginal_params}.
A parameter prior is a probability distribution which
we know \textit{a priori} might be true for given possible values of a parameter in a model.
In implementations of the Bayesian approach for detecting DLAs in quasar spectra \citep{mfDLA, mfDLA16},
the prior distribution for column density
was learned from previous DLA catalogues, and they used a uniform absorber red-shift prior distribution.

One of the input parameters in the Voigt profile\footnote{
See \texttt{voigt\_IP.c} in \href{https://github.com/rezamonadi/GaussianProcessCIV}{https://github.com/rezamonadi/GaussianProcessCIV}}
is the absorber column
density, $N_{\civ}$.
Following \cite{romanDLA} and \cite{mfDLA}, we need to sample a
column density distribution to perform the integral in Equation \ref{eq:marginal_params}
and obtain the model likelihood.

The column density range detected by \citet{C13} was $\log \nciv \approx 13$ to $>15.8$.
After some experimentation we chose
 a slightly larger range: $12.5 < \log_{10} \nciv < 16.1$, which maximised the performance on our validation set.
We probed larger column densities than PM catalogue because: first, their column densities are often lower limits as
they used the apparent optical depth method \citep{AODM} and a lot of the absorption systems
were saturated.  Second, the larger size of SDSS DR12
gives us a longer survey pathlength  which increases our chances of finding the exponentially rare strong systems.
We searched for lower column densities than the PM catalogue
since our catalogue could potentially be more sensitive to weaker absorbers (see Figure \ref{fig:colDenCompar}
for the posterior distribution of column densities). We thus
used a mixture probability density function consisting of:
(1) the $\nciv$ probability density function (obtained by kernel density estimation)
from the reported values in the PM catalogue  and
(2) a uniform probability density function in the same range.
We have also confirmed that our column density prior sample reproduces a
rest equivalent width ($W_{r,1548}$) distribution in reasonable agreement with the PM catalogue for the 1548~\AA\ line.
     
We also need a prior for the Doppler velocity dispersion, $\sigma_v$.
The typical temperature for the intergalactic medium is $\sim 10^4\!-\!10^5$
K, which gives a $\sigma_v \sim 2.6\!-\!8.3$ \kms\ for \civ.
However, at the low resolution of the SDSS spectra ($\sim 150$ \kms), it is impossible
to detect an absorption line with this velocity dispersion.
Fortunately, \civ\ absorbers cluster \citep{Boksenberg2003} and blend into a broader absorption profile with larger effective
$\sciv$. By experimenting with different ranges for $\sciv$
we chose lower and upper bounds for $\sciv$ to be 35~\kms and 115~\kms, respectively.
Note that we used the hold-out sample described in Section \ref{sec:data} and tried different $\sciv$\ ranges
to obtain the highest classification performance and  purity/completeness (see Section \ref{sec:validation}).
Moreover, we ensured that this range enables our process to be sensitive to similar rest equivalent widths as the PM catalogue.

We imposed a uniform prior distribution on the absorber redshift, $\zciv$.
The lower limit is the redshift at which the 1548~\AA\ line is observed at
$1310(1+\zqso)$,\footnote{To avoid possible confusion with any \ion{O}{I}, \ion{Si}{II} absorption pairs, see \cite{C13}.} or the
blue end of our input spectrum whichever is larger. Therefore:
\begin{equation}
1+ z_{\rm min} = \max\left[\frac{\min(\lambda_{\rm obs})}{1548}, \frac{1310(1+\zqso)}{1548}\right].
\label{eq:zmin}
\end{equation}
We also require a small velocity separation between the absorber and the quasar, to ensure that we are not
finding intrinsic \civ\ absorbers around the host galaxy of the quasar:
\begin{equation}
z_{\rm max} = \zqso - \frac{\delta {\rm v}}{c}(1+\zqso).
\label{eq:zmax}
\end{equation}
We considered $\delta {\rm v} = 1000$ to $5000$ \kms, and achieved the best validation
performance when $\delta {\rm v} =3000$ \kms, which matches the minimum
velocity separation between quasar and absorbers in the PM catalogue.

We assumed that $\nciv$ and $\sciv$ are independent from $\zqso$, although
$\zciv$ depends on $\zqso$ as described in Equation \ref{eq:zmin} and Equation \ref{eq:zmax}.
We calculated the marginalised model likelihood by integrating the absorption-model priors $\model_{D\slash S}$\footnote{Either the doublet model or the singlet model.} as:
\begin{equation}
P(\theta \vert \zqso)  \propto P(\zciv \vert \zqso)P(\nciv)P(\sciv).
\label{eq:pr_model}
\end{equation}
Then we performed the integral for our absorption models, $\model_D$ and $\model_S$, in Equation \ref{eq:marginal_params}:
\begin{equation}
    P(\Data \vert \zqso) \propto
    \int P(\vec{y} \vert \theta, \zqso)
    P(\theta \vert \zqso)\dd\theta.
\label{eq:model_evidence}
\end{equation}
However, Equation \ref{eq:model_evidence} is intractable, so we approximated it
with a quasi-Monte Carlo method. This method selected $10,000$
samples of $\{\nciv, \sciv, \zciv\}$ at which to calculate the model
likelihood. The samples were drawn from a Halton sequence to ensure an approximately uniform spatial distribution. We approximate the model evidence by the sample mean:
\begin{equation}
  P(\Data \vert \model_{D\slash S}, \zqso) \simeq
  \frac{1}{N} \sum_{i=1}^{N} P(\Data \vert \theta_i, \zqso, \model_{D\slash S}).
\label{eq:model_evidence_qmc}
\end{equation}
We integrated out the parameters, $\theta = \{\zciv,\nciv, \sciv\}$, with
a given parameter prior $P(\theta \vert \zqso, \model_{D/S})$.
We use 10,000 samples: lower sample sizes under-sample the likelihood function, while larger sample sizes
cause the code to run slower.
We considered 10,000--50,000 %
samples in the validation phase and found that increasing the number of samples did not significantly improve
the validation performance. Note that using more samples increases the run-time cost of processing a quasar.
In calculating the model evidence for the singlet model, $\model_S$,
we used a single component Voigt profile centred on 1548~\AA\ (Equation \ref{eq:model_s}) while for calculating the
model evidence for the doublet model, $\model_D$, we use a double component Voigt
profile centred at 1550~\AA\ and 1548~\AA\ (Equation \ref{eq:model_D}).
We used the same parameter priors for both the singlet and doublet models for simplicity.

\subsection{Multiple absorber search}
\label{sec:multiciv}

In this paper, instead of reporting probabilities for multiple \civ\ absorbers as
\cite{mfDLA} did for DLAs, 
we
simplified and reported the probability that there is an absorber at a given redshift.
For example, the posterior probability for the $k=3$ model in \cite{mfDLA} does not indicate which of these
three absorbers in $\model_{DLA(3)}$ is most probable, instead reporting the probability that a given spectrum contains some combination of three absorbers.

\begin{figure}
\includegraphics[width=\linewidth]{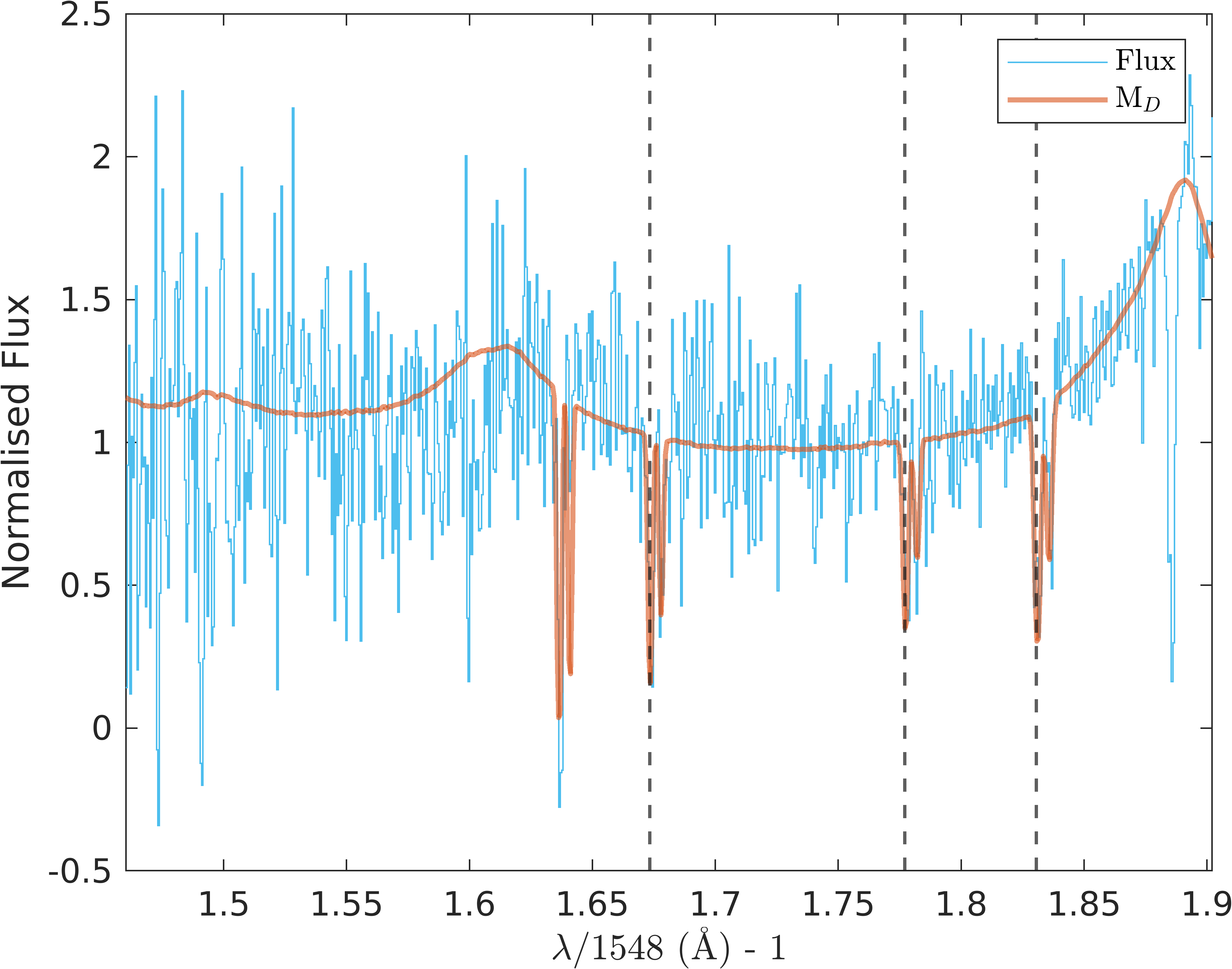}
\caption{Example SDSS DR7 spectrum with QSO-ID: 51608-0267-264 and $\zqso = 1.89$.
Both PM and our pipeline find three absorbers between $\zciv = 1.65$--1.85.
We also find an absorber at $\zciv=1.489$ (probability 92\%) that
was not detected by PM, due to noise in this part of the spectrum
(specifically, the 1550 line was not automatically detected with
their parameters, thus the doublet was not visually inspected).
The probabilities that our pipeline provides for the existence of the first, second, third,
and forth \civ\ absorber are
$P(\civ)=[1.00, 1.00, 1.00, 0.92]$, respectively, 
our maximum a posteriori absorber redshift values are
$\zciv=[1.829, 1.672, 1.775, 1.489]$, and our rest equivalent widths from Voigt
profile integration (see Equation \ref{eq:rewgp}) are ${W_{r,1548}^{\rm GP}}=[ 1.37, 0.87, 0.90, 0.79]$~\AA.
In the PM-catalogue the absorber redshifts are
$z_{\rm PM}=[ 1.831, 1.673, 1.777]$ with corresponding
${W_{r,1548}^{\rm PM}}=[1.21\pm0.18, 1.40\pm0.20, 0.94\pm0.19]$~\AA}
\label{fig:PM1GP1}
\end{figure}

Here, we wish to find multiple absorbers in a spectrum. We proceed iteratively, noting that
at any point the best-fit may be a singlet or a doublet, and mask out the most likely absorber
each time. We mask
350 \kms around 1548~\AA$({\rm MAP}(\zciv)+1)$ and 350 \kms around
1550~\AA$({\rm MAP}(\zciv)+1)$, where MAP$(\zciv)$ is the maximum \textit{a posteriori} value for  $\zciv$. For single-line absorbers, we mask 350~\kms around 1548~\AA, again at MAP$(\zciv)$. Our procedure is as follows:
\begin{enumerate}
\item[1)] Fit our three models $\model_{N/S/D}$ on an observed spectrum. \label{step:i}
\item[2)] If $\model_N$ (the null, \civ-free, model) has the highest posterior for any search,
there is no \civ\ absorption in the given spectrum. Stop any further searches. Otherwise go to step 3.
\item[3)] If either $\model_S$ or $\model_D$ has the highest posterior, 
mask the spectral region around the most probable absorption profile.
Return to step 1 to search for subsequent absorbers if no more than seven searches previously
have been done. Otherwise stop any further searches.
\end{enumerate}
Figure \ref{fig:PM1GP1}
shows an example quasar spectrum (SDSS DR7 QSO-ID: 51608-0267-264 and $\zqso = 1.89$)
within
which both the PM and GP pipelines find three absorbers.
Moreover, the GP pipeline finds an absorber at $\zciv=1.489$ (probability 92\%) that
was not detected by PM, due to noise in this part of the spectrum.
Specifically, the 1550 line was not automatically detected with
their parameters, thus the doublet was not visually inspected.

\section{Validation}
\label{sec:validation}

For validation, we trained a \civ-free model, $\model_{N}$, on a reduced training set of 95\% of the
inspected spectra in the PM catalogue \citet{C13}.
We then \emph{validated} our algorithm with the remaining 5\% of the
inspected (1301) spectra in the PM catalogue to
check the agreement between the PM catalogue and our method.
Note that when we applied our algorithm to DR12 spectra, we re-trained our model
using all SDSS DR7 spectra inspected in the PM catalogue without
a reliable \civ~absorber.

Our model is compared to the \civ\ absorbers as rated in the PM catalogue. \citealt{C13} rated
their automatically detected \civ\ candidates
from 0 (definitely not \civ), 1, 2, and 3 (definitely \civ), thus providing a
rough estimate of confidence in an absorber. Absorbers with a ranking $\ge 2$ are considered
real \civ\ absorbers in the PM catalogue.
We construct a  ``ground truth''  sample of the PM \civ\ with rating $\ge2$.
Within a spectrum, we enforce that our GP-detected absorber is within 350 \kms\ of a
PM-detected system to be considered as a ``match'' between catalogues (see Figure \ref{fig:PM1GP1} for examples of matched absorbers);
this cutoff is roughly  $3\times \textrm{max}(\sciv)$ (where $\sciv$\ is measured by the GP),\footnote{For reference, in \cite{C13}, \civ\ absorbers were grouped into a single system if they were within 250~\kms\ of each other.} which ensures we are not
detecting a complex\slash blended 
system in two successive iterations (see Section \ref{sec:multiciv}).
Moreover, we obtained a better
purity\slash completeness (see Section \ref{sec:purity_completeness}) with a 350~\kms\ masking window.

\subsection{Velocity separation}
\label{sec:velocity}
\begin{figure}
\includegraphics[width = \linewidth]{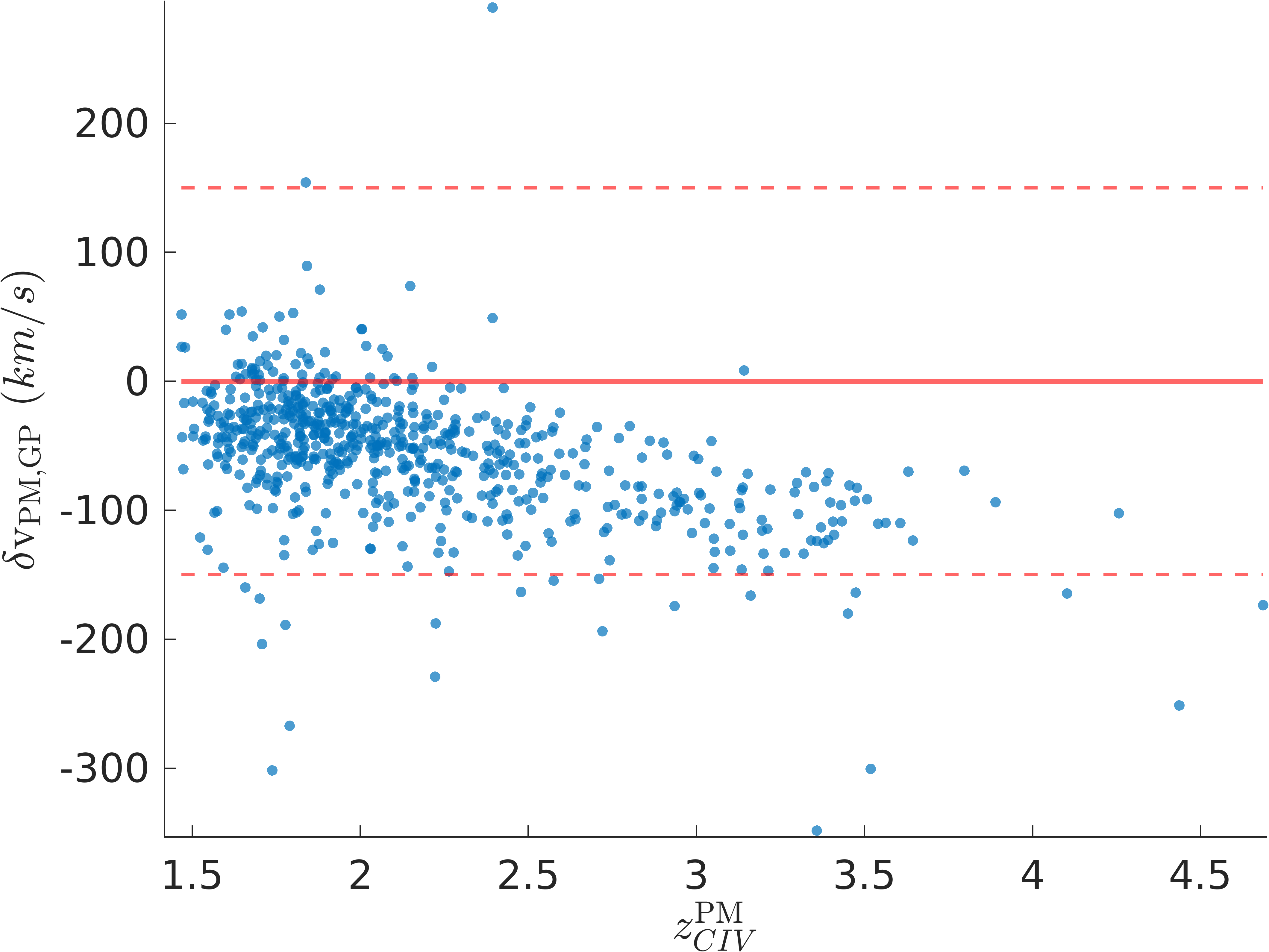}
\caption{Velocity difference between  the detected absorbers in the
GP pipeline with ${\rm P}(\model_D)\ge0.95$ in the validation set and the absorbers in the PM catalogue. Only absorber pairs closer than 350~\kms are shown. The thick red line shows $\delta {\rm v}_{\rm PM,GP}=0$ and
the dashed lines are  $\delta {\rm v}_{\rm PM,GP}=\pm150$~\kms (the SDSS spectral resolution).
The median offset is $\delta {\rm v}_{\rm PM,GP}^{\rm med}\approx$-50~\kms, which is less than an SDSS pixel (69~\kms).}
\label{fig:dz}
\end{figure}

The velocity separation between absorbers detected in both the GP and PM catalogues is:
\begin{equation}
\label{eq:dvGPPM}
\delta {\rm v}_{\rm PM,GP} = \frac{\zciv^{\rm PM}-\zciv^{\rm GP}}{1+\zciv^{\rm PM}}c.
\end{equation}
Figure \ref{fig:dz} shows that absorber redshifts obtained by our pipeline in the validation set are almost always consistent with
the PM catalogue at the level of the SDSS spectral
resolution, i.e., $\vert\delta {\rm v}_{\rm GP,PM}\vert \lesssim 150$\,\kms.
Very few points in Figure \ref{fig:dz} lie outside of the $\pm$150~\kms horizontal lines.
Our pipeline produces $\zciv$\ on average slightly greater$\slash$ redder than
the PM catalogue, with a median offset $\delta {\rm v}_{\rm PM,GP}^{\rm med}$$\approx$-50~\kms.
This is not a significant difference; by comparison an SDSS pixel is 69~\kms.
In the PM catalogue, the  redshift of the 1548~\AA\ line was sometimes underestimated, as
redshift estimation is weighted by flux-centered centroid   where 
lower redshift wavelength pixels than 1548\AA\ have higher flux values. This can be more common
at higher redshifts.

\begin{figure}
\includegraphics[width=\linewidth]{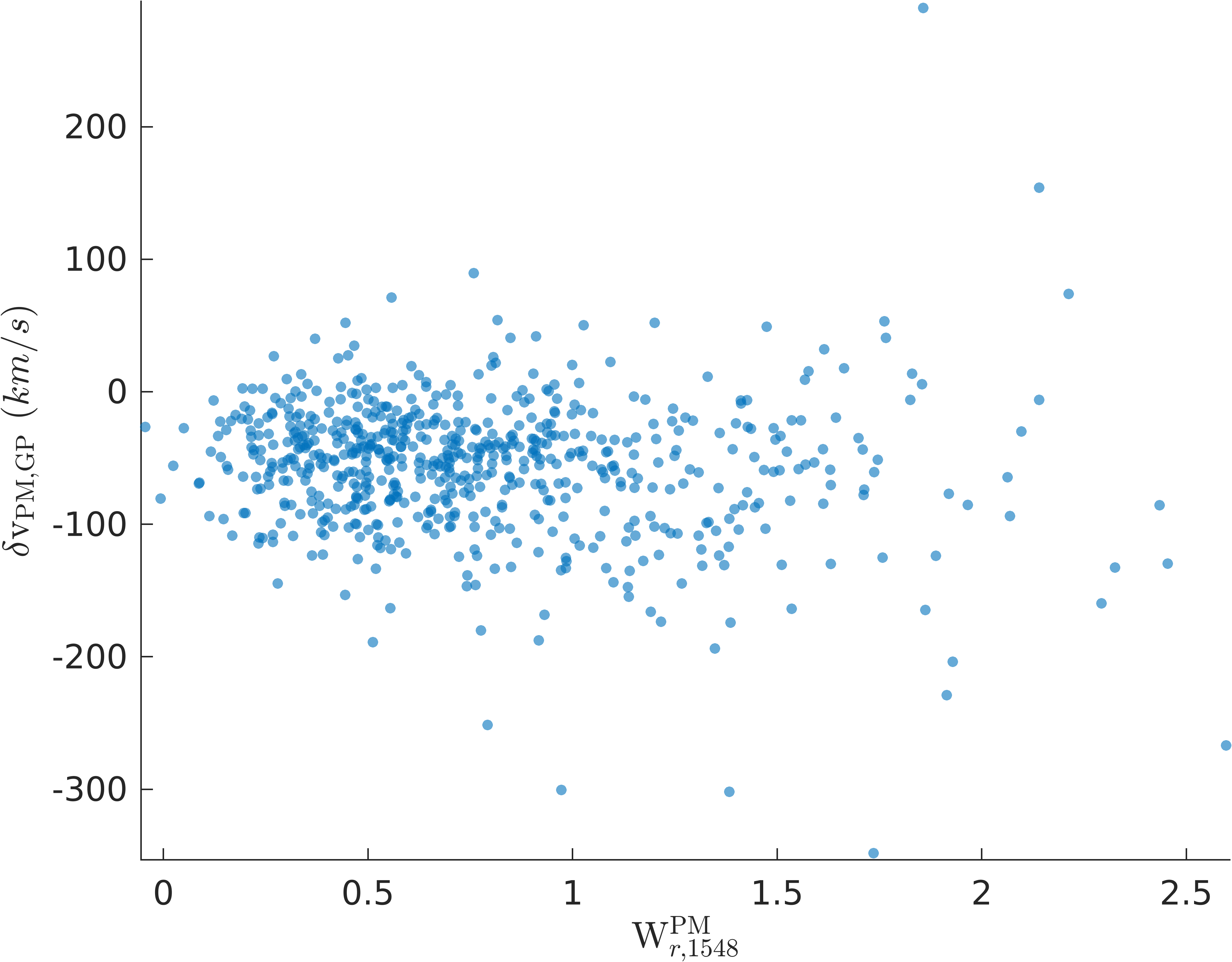}
\caption{Velocity separation (Equation \ref{eq:dvGPPM}) between GP and PM detected \civ\ absorption
systems is shown versus the reported rest equivalent width values for 1548 \AA\ in the PM
catalogue (${\rm W}_{r,1548}^{\rm PM}$).
There is no correlation between the velocity separation and the strength of detected absorbers. }
\label{fig:dvGPPM-EW}
\end{figure}

We visually inspected the 9 spectra in our validation set of 1301 spectra with  $\delta {\rm v}_{\rm PM,GP}\ge50$~\kms:
most of them were in a complex/blend system and some of them were close to the QSO where the GP continuum was not perfect. We also investigated the 14 spectra in the validation set that
show  $\delta {\rm v}_{\rm PM,GP}\le-150$~\kms: most of them belong to a complex system or even a mini-BAL system.
In some cases the GP continuum fit is not good. As a reference, we investigated 17 spectra
with  $\delta {\rm v}_{\rm PM,GP}\sim-50$~\kms: these spectra are usually high SNR and/or the GP continuum
fit is very good, especially around the detected absorption system. Moreover, there is no significant
correlation between the strength of the absorber systems and PM-GP velocity separation
(Equation \ref{eq:dvGPPM}) as shown by Figure \ref{fig:dvGPPM-EW}.

\subsection{Receiver Operator Characteristic (ROC) curve}
\label{sec:roc}
\begin{figure}
\centering
\includegraphics[width=\linewidth]{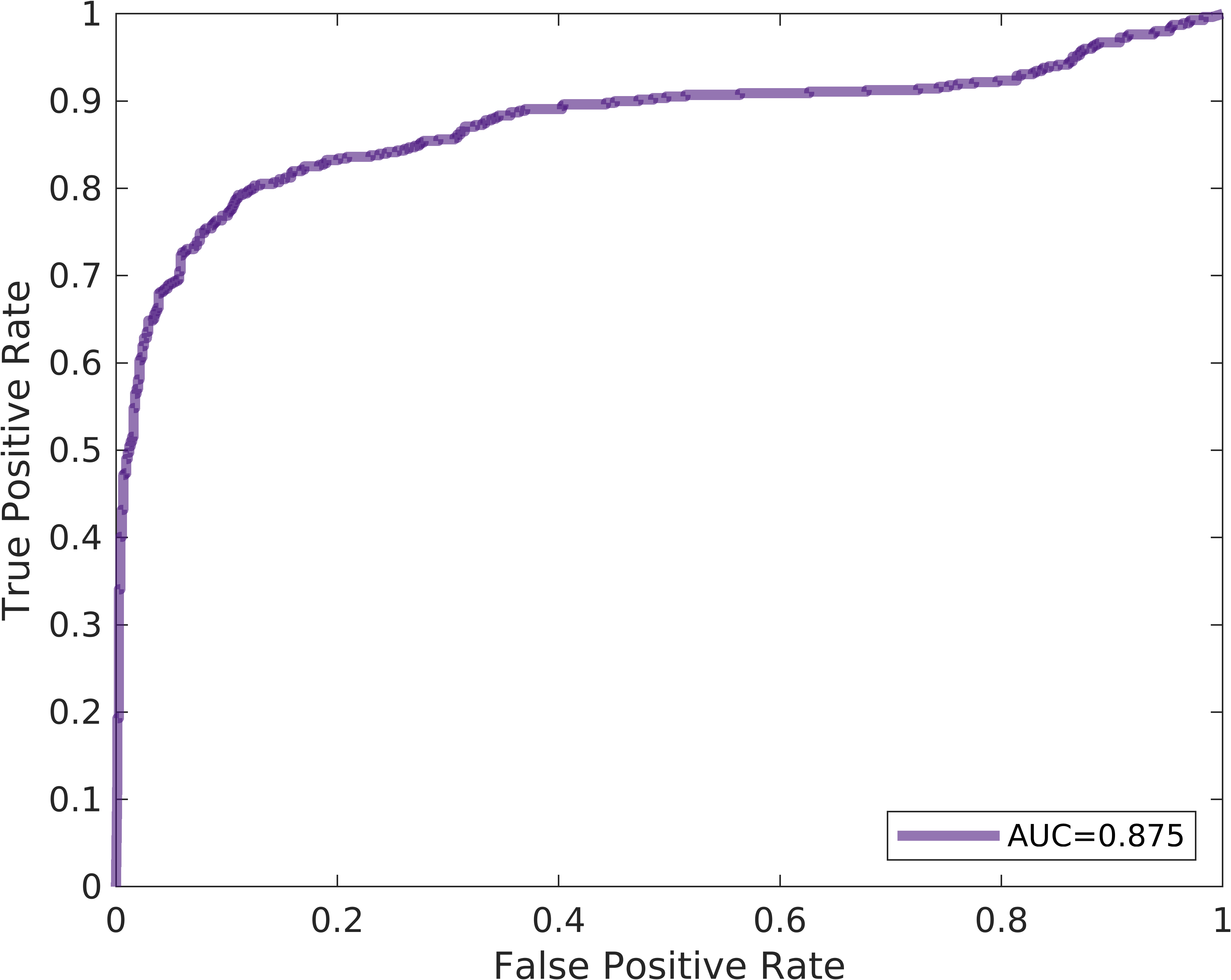}
\caption{Receiver Operator Characteristic (ROC) curve for our DR7 validation.
True Positive Rate  is plotted versus False Positive Rate. True positives are \civ\ systems in our catalogue
at least 350~\kms apart from an absorber in the  PM catalogue with ranking  $\ge$2 given
  any ${\rm P}(\model_D)$ threshold between 0 and 1.
False positives are those absorbers in our catalogue that do not have any
matching absorber in the PM catalogue;
though they may be real \civ\ absorbers (see Figure \ref{fig:PM1GP1}).
Above a relatively small
False Positive Rate ($\sim 0.2$), our algorithm procedure obtains True
  Positive Rate above 80\% and, hence, is a successful
way to identify \civ\ absorbers.  The area
under the ROC curve ($\textsc{AUC}$) is a quantitative metric for the equality of
the GP algorithm;
we get $\textsc{AUC} = 0.87$, a reasonable value compared to an ideal classification that gives
$\textsc{AUC} = 1.00$.
}
\label{fig:ROC}
\end{figure}

We use the Receiver Operator Characteristic (ROC) curve (Figure \ref{fig:ROC}), which is
the true positive rate versus false positive rate
for any classification threshold: $0\le P(\model_{D}) \le 1$ to obtain a 
score out of 1 for the performance of our classification (no \civ absorber versus \civ absorbers).
The Y-axis of the ROC curve in Figure \ref{fig:ROC}, the true positive rate, is the ratio
of the number of \civ\ absorbers in our catalogue to the total number of
of absorbers in the PM catalogue with a ranking $\ge2$. \civ\ absorbers in our catalogue are defined to be those with posterior probability greater than a threshold, ${\rm P}(\model_D)$, between $0$ and $1$. They must also be less than $350$~\kms apart from an absorber in the PM catalogue with a ranking$\ge 2$. The X-axis of the ROC curve in Figure \ref{fig:ROC}, the false positive rate, is the ratio of  \civ\ absorbers in our catalogue
that do not have any matching absorber with ranking $\ge 2$ in the
PM catalogue (given any ${\rm P}(\model_D)$ threshold between 0 and 1) to the total number of absorbers in the PM catalogue with a ranking $\ge 2$.

A higher classification performance (i.e. in each search run over a spectrum we classify it
as \civ-free or having a \civ\ absorber) is reflected in a larger area under the   curve (\textsc{AUC}) for the ROC curve.
We obtain a quite reasonable $\mbox{\textsc{AUC}}=0.87$. 
Note that here ``true positive'' refers to a PM \civ\ absorber recovered by the GP
algorithm in the training set, and ``false positive'' is a GP \civ\ absorber not in the PM catalogue.
However, as seen in Figure \ref{fig:PM1GP1}, the GP procedure \textit{can} find real\slash true \civ\ absorption
not identified in the PM survey; hence, ``false positives'' may be better considered ``GP unique''. This also means that
the classification performance ($\mbox{\textsc{AUC}}=0.87$) we obtained here might underestimate the true performance.

\subsection{Purity and Completeness}
\label{sec:purity_completeness}
\begin{figure}
\includegraphics[width=\linewidth]{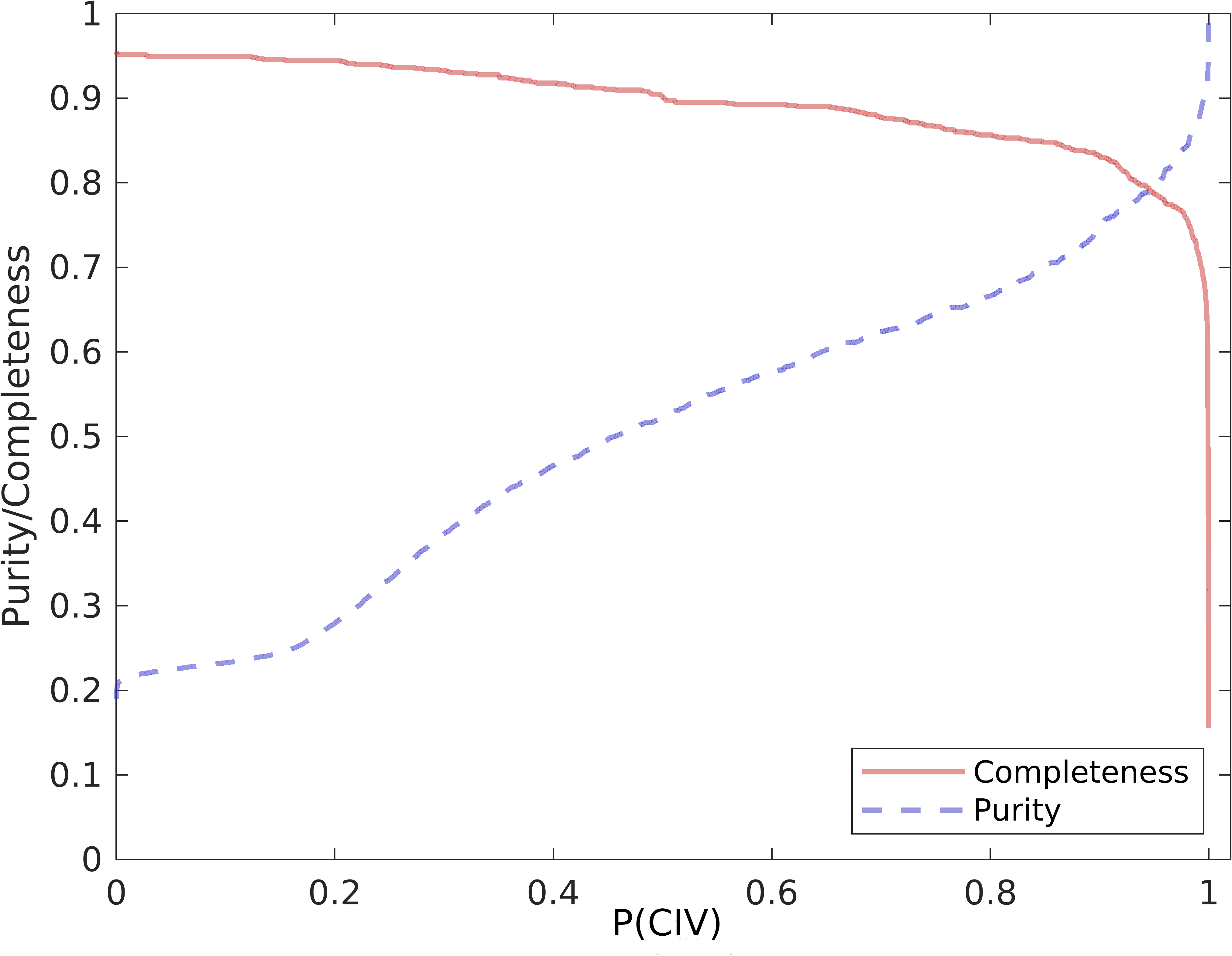}
\caption{Purity (Equation \ref{eq:purity}) and completeness (Equation \ref{eq:completeness}) of the GP catalogue compared to the PM catalogue for different \civ\
posterior probability (Equation \ref{eq:model_selection}) thresholds.
The maximum allowed velocity separation between our catalogue and the PM catalogue absorbers is 350 \kms.
The intersection of the purity
 (dashed blue curve) and completeness (solid red curve) at a threshold of $\sim 95\%$ gives us a balanced
 purity/completeness of $\sim 80\%$.}
\label{fig:threshold}
\end{figure}

We  assessed our algorithm's performance by comparing individual absorption systems. We can compare our
GP catalogue for various \civ\ posterior probabilities
to the `ground truth' sample of the PM catalogue.
We define the purity of our GP catalogue as the fraction of the GP catalogue also in the PM catalogue:
\begin{equation}
  \label{eq:purity}
\mathrm{Purity} = \frac{\mathrm{GP} \cap \mathrm{PM}}{\mathrm{GP}}.
\end{equation}
The completeness is the fraction of the PM catalogue also in the GP catalogue:
\begin{equation}
  \label{eq:completeness}
\mathrm{Completeness} = \frac{\mathrm{GP} \cap \mathrm{PM}}{\mathrm{PM}}.
\end{equation}
Figure \ref{fig:threshold} shows completeness and purity as a function of threshold value.
One should choose a threshold that gives the best possible
combination of purity and completeness, around the point where the curves intersect.
We thus choose a threshold of 95\%, which Figure \ref{fig:threshold} shows produces purity and completeness of $\sim 80\%$ in a roughly equal balance.
However, our catalogue reports posterior probabilities, so the user may choose a different threshold as
desired for their application. 

One of the strengths of our catalogue is the freedom that the user has for choosing the absorbers 
based on the \civ\ absorption model posterior probability P($\model_D$). The user can sacrifice the
purity of absorbers for their completeness or vice versa.
By sacrificing the purity, we are accepting absorbers with lower P($\model_D$). This will
increase the number of accepted absorbers. However, it increases the chance of 
misidentifying some absorption features as \civ\ ones. By sacrificing   the 
completeness, we are restricting our catalogue to absorbers with higher P($\model_D$).
This will lower the number of accepted absorbers but will boost the purity of our catalog. 

\subsection{Rest equivalent width comparison}
\label{sec:ValidREW}
We can evaluate our algorithm by comparing 1548~\AA\ rest equivalent
width ${\rm W}_r^{\rm GP, flux}$ between the GP and PM catalogues. ${\rm W}_r^{\rm GP, flux}$ is obtained by integrating
the normalised flux deficit from our GP continuum ($\model_N$)
in a wavelength integration window corresponding to $4\times\sciv$ around the maximum \textit{a posteriori} $\zciv$ for the 1548~\AA\ line.
We impose that the flux integration window does not
exceed the midpoint of the 1550~\AA\ and 1548~\AA\ lines.

\begin{figure}
  \includegraphics[width=\linewidth]{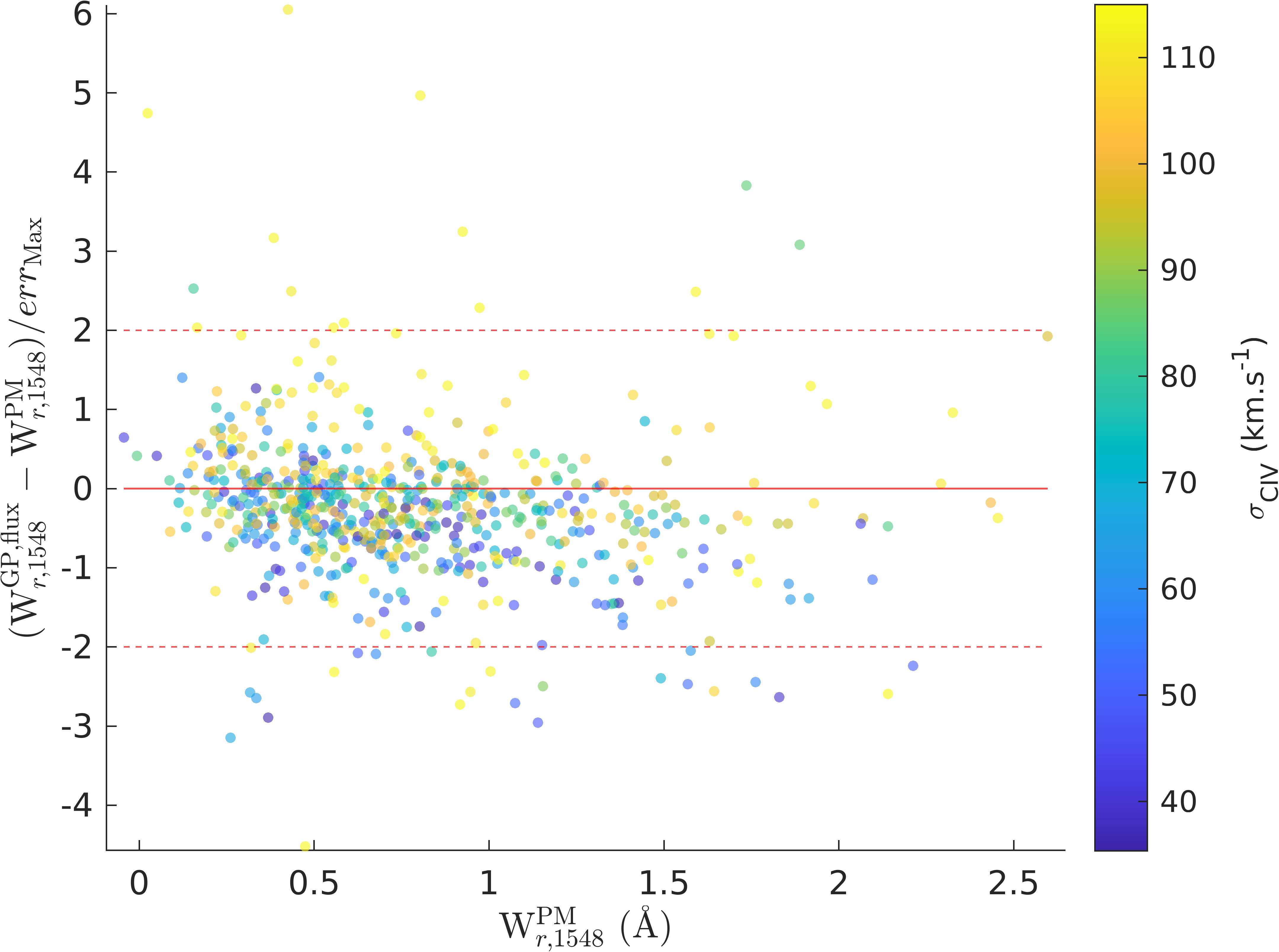}
  \caption{The ratio of the difference between rest equivalent width from our pipeline with boxcar flux summation
  (${\rm W}_{r,1548}^{\rm GP, flux}$) and rest equivalent width from the PM catalogue (${\rm W}_{r,1548}^{\rm PM}$) to the total error (see Equation \ref{eq:totalErr})
  from the PM catalogue and our pipeline for $W_{r,1548}$. The data points here are  those
  absorption systems in the validation set where
  our pipeline reports an absorber with ${\rm P}(\model_D)\ge 0.95$ and
  for which there is an absorber with ranking $\ge2$ in the PM catalogue at
  a redshift offset less than 350 \kms (GP \& PM in Section \ref{sec:ValidREW}).
  As the colour bar shows, there is a trend towards larger maximum \textit{a posteriori} $\sciv$\ when the
  GP rest equivalent width is larger than the rest equivalent width from the PM catalogue.}
  \label{fig:errREWPMGP}
\end{figure}

Figure \ref{fig:errREWPMGP} shows the difference ratio
in our validation set between the rest equivalent width
of the 1548\AA~line
in the GP catalogue and the PM catalogue, scaled by the maximum error (because the rest equivalent width errors from the PM and GP catalogues are highly correlated):
\begin{equation*}
\frac{{\rm W}_{r,1548}^{\rm GP, flux} - {\rm W}_{r,1548}^{\rm PM}}{{ \mathrm{err}}_{\rm max}}.
\end{equation*}
The maximum error, $\mathrm{err}_{\rm max}$, is obtained by comparing the rest equivalent width error from the GP pipeline to that from the PM catalogue:
\begin{equation}
\mathrm{err}_{\rm max} = \max\{\mathrm{err}({\rm W}_{r,1548}^{\rm GP, flux}), \mathrm{err}({\rm W}_{r,1548}^{\rm PM})\}.
\label{eq:totalErr}
\end{equation}
We obtained $\mathrm{err} ({\rm W}_{r,1548}^{\rm GP, flux})$  by considering the observed noise in each pixel included in the integration window described above.
%

Around 94\% 
of the data points in Figure \ref{fig:errREWPMGP} have
$\vert \left({\rm W}_{r,1548}^{\rm GP, flux} - {\rm W}_{r,1548}^{\rm PM}\right) / \mathrm{err}_{\rm max} \vert \le 2$, which shows a reasonable
consistency between the GP and PM rest equivalent widths.
We visually inspected all of the 21 absorbers with
$\left({\rm W}_{r,1548}^{\rm GP, flux} - {\rm W}_{r,1548}^{\rm PM}\right) / \mathrm{err}_{\rm max} <-2$:
they mostly have continuum issues and a low GP continuum. For
QSO 53886-1823-377,  a triplet\footnote{When a lower redshift absorber's 1550~\AA\
  line blends with the 1548~\AA\ line of the higher redshift absorber.} \civ\ system
  at $\zciv= 1.838$ caused  a lower 1548 rest equivalent width in the GP catalogue than in the
  PM catalogue. In the spectrum of QSO 53083-1757-529  our algorithm finds
  an absorber at the end of the spectrum, which also yields a lower rest equivalent width when compared to the PM catalogue.
  Looking at 13 absorbers
  with $\left({\rm W}_{r,1548}^{\rm GP, flux} - {\rm W}_{r,1548}^{\rm PM}\right)/ \mathrm{err}_{\rm max} >2$,
we realised that most of these absorbers belong to a triplet \civ\ system.
As a reference  we also checked absorbers with $\left\vert {\rm W}_{r,1548}^{\rm GP, flux} - {\rm W}_{r,1548}^{\rm PM}\right\vert / \mathrm{err}_{\rm max} <0.025$:
these spectra were mostly high SNR and the GP continuum fit the observed quasar very well.

The colour bar in Figure \ref{fig:errREWPMGP} shows the \textit{maximum a posteriori} $\sciv$\
that the GP algorithm produces for each absorber.
There is a correlation between larger \textit{maximum a posteriori} $\sciv$\ and absorbers where the GP rest equivalent width, ${\rm W}_{r,1548}^{\rm GP,flux}$,  is larger than the PM rest equivalent width, ${\rm W}_{r,1548}^{\rm PM}$. We visually inspected these systems and found that many of them are triplet or mini-BAL systems, for which the GP is more likely to give a large $\sciv$.

The difference in rest equivalent widths of the 1550~\AA~line between the GP and PM catalogues behaves similarly. We find that 518 (86\%) of GP absorbers with a
PM absorber system at a redshift offset less than 350~\kms\ away showed
$\left\vert{\rm W}_{r,1500}^{\rm GP,flux} - {\rm W}_{r,1500}^{\rm PM}\right\vert/\mathrm{err}_{\rm max} \le 2$.
The $1550$~\AA\ line is weaker than the $1548$ \AA~line, leading to a generally lower detection significance. However, for strong absorbers it is useful because it is less saturated.

The GP pipeline finds 822 absorbers in the validation set spectra with
P$(\model_D)\ge95\%$. In the PM catalogue the validation set spectra contain 829 absorbers with a ranking $\ge 2$.
We can divide these absorbers into four different categories
with the following statistics:
\begin{enumerate}
  \item PM \& GP: absorbers with a ranking$\ge 2$ in
  the PM catalogue,  ${\rm P}(\model_D)\ge0.95$ in the GP catalogue, and $\delta {\rm v}_{\rm PM, GP}\le 350$~\kms.
   This category contains 647 absorbers, $\sim78\%$ of the PM absorbers
   and  $\sim79\%$ of the GP absorbers among the 1301 spectra in the  validation set.
  \item GP only: absorbers with ${\rm P}(\model_D)\ge0.95$ but no absorber
  in the PM catalogue with ranking $\ge 2$ and a velocity offset less than 350 \kms.
  This category includes 175 absorbers (21\% of the GP absorbers in the validation set). Some of these absorbers are true \civ~which fell beneath the sensitivity of the candidate search in \cite{C13}, and some are other doublet lines which our GP model has incorrectly classified as \civ.
  \item GP uncertain: absorbers with a ranking $\ge2$ in the PM catalogue, and an absorber from the GP catalogue with a velocity offset less than 350 \kms\ but P$(\model_D)<95\%$. There are 142 of these absorbers, $\sim 17\%$ of the PM absorbers in the validation set spectra.
   Note that 85  ($\sim 60\%$) of these GP uncertain absorbers have P$(\model_D)\ge 50\%$ in the GP catalogue and that the authors of the PM catalogue resolved ambiguous absorbers by inspecting other metal lines from the same system.
    \item PM only: $40$ (4.8\%) of 829 absorbers with ranking $\ge2$ in the PM catalogue validation set had no GP absorber candidates within $350$ \kms in the GP catalogue.
    The GP pipeline thus misses these absorbers in its successive searches of the validation set spectra. Two absorbers were assigned to this category because there were two PM absorbers in the spectrum closer than 350 \kms\ to each other. The GP catalogue found one, and the region containing the second was masked. Note that these absorbers are the reason why Figure \ref{fig:threshold} does not show a completeness of $1$, even with a threshold of P$(\model_D)=0$.
   \end{enumerate}

\begin{figure}
\includegraphics[width=\linewidth]{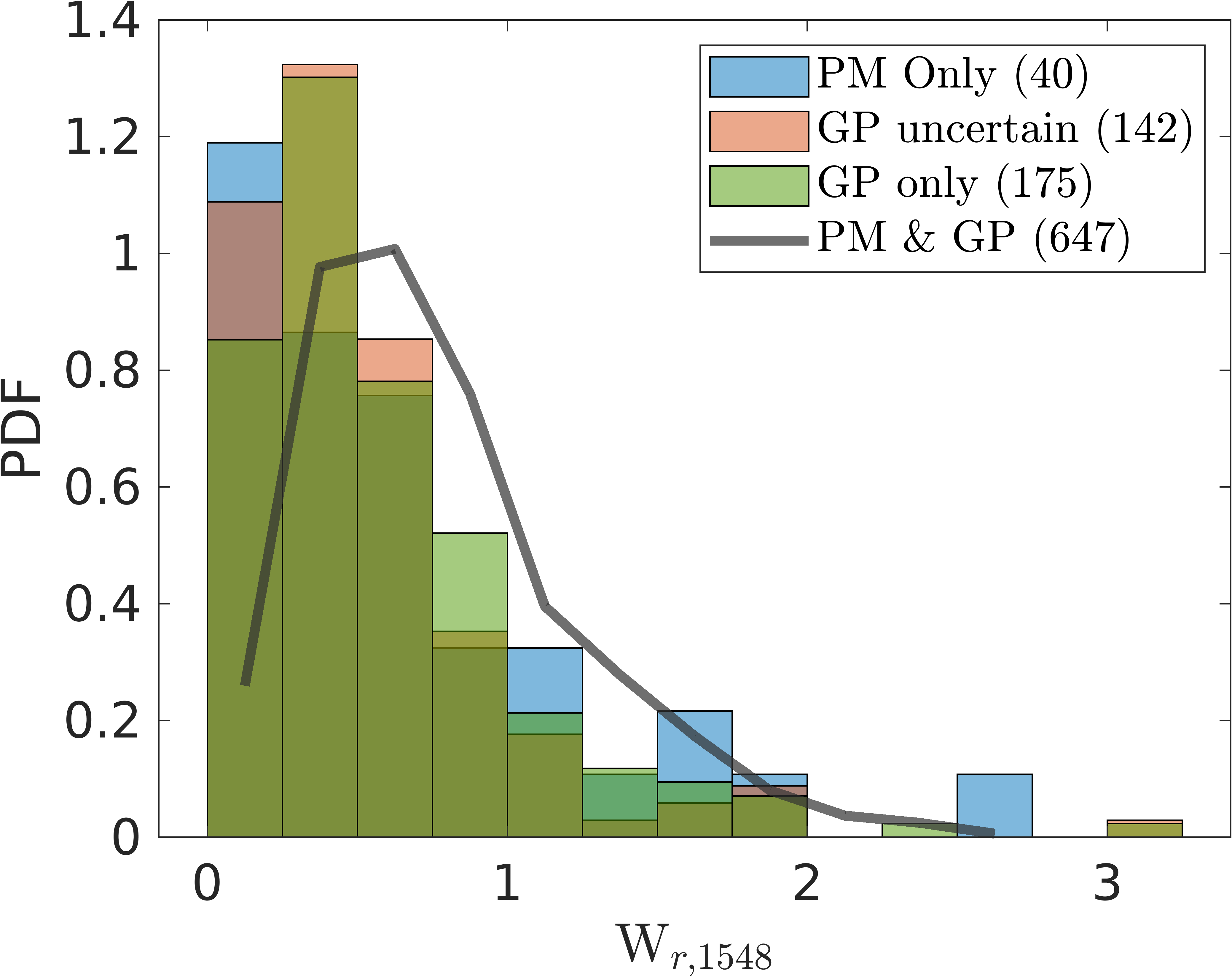}
\caption{Distribution of W$_{r,1548}$ for absorbers in four categories described in
Section \ref{sec:ValidREW}:
detected in both the GP and PM catalogues
(thick black line), in the GP uncertain (brown), in GP only (green), in the PM catalogue only (blue).
The rest equivalent width distribution is similar for all categories. There are some strong absorbers with (W$_{r,1548}>1.2$\AA) classified as ``PM only''. Visual inspection of
the spectra of these systems indicates that they are part of a triplet/complex absorber or a broad mini-BAL system.
}
\label{fig:REWDR7}
\end{figure}

  Figure \ref{fig:REWDR7} shows the distribution of rest equivalent widths for
  40 PM only absorbers, 142 GP uncertain absorbers, 175 GP only absorbers, and 647 PM \& GP absorbers
  in the four categories described above.  
  Figure \ref{fig:REWDR7} also demonstrates
  that weak absorbers ($W_{r,1548}$<0.3\AA) 
  are detected less often than other categories.
  Considering that there are intrinsically more weak absorbers in the population, it is likely that
  each algorithm (i.e GP and PM) captures a different subset of weak absorbers. However, for
  stronger absorbers, the four categories are more or less consistent considering the uncertainties in
  measuring rest equivalent width.

%
There are 17 strong absorbers (W$_{r,1548}\ge1.2$\AA) in the PM only or GP uncertain categories.
11 of these absorbers are triplet/mini-BAL systems where the GP pipeline gives P$(\model_S) \sim P(\model_D)$. The complex shape of these absorption systems are not a good match to either model, so the GP pipeline is not able to distinguish between them.
Two absorbers have P$(\model_D)\ge0.95$ but a velocity separation more than 350~\kms. One of these is also close to a complex absorber system. Five absorbers are detected by the GP catalogue as a singlet, and so have $P(\model_S) \gg P(\model_D)$.

 Thus most missed strong absorbers are caused either by \civ~triplets or by the GP pipeline preferring a singlet fit to a doublet in cases where the doublet structure is not well resolved.
Note that when conducting visual inspection of \civ\ absorbers, an observer
may resolve ambiguous lines using information from other metal line transitions associated with the same system, whereas our GP pipeline uses only information from the \civ~transition.

\subsection {Example absorbers}
\label{sec:examples}

\begin{figure}
\includegraphics[width=\linewidth]{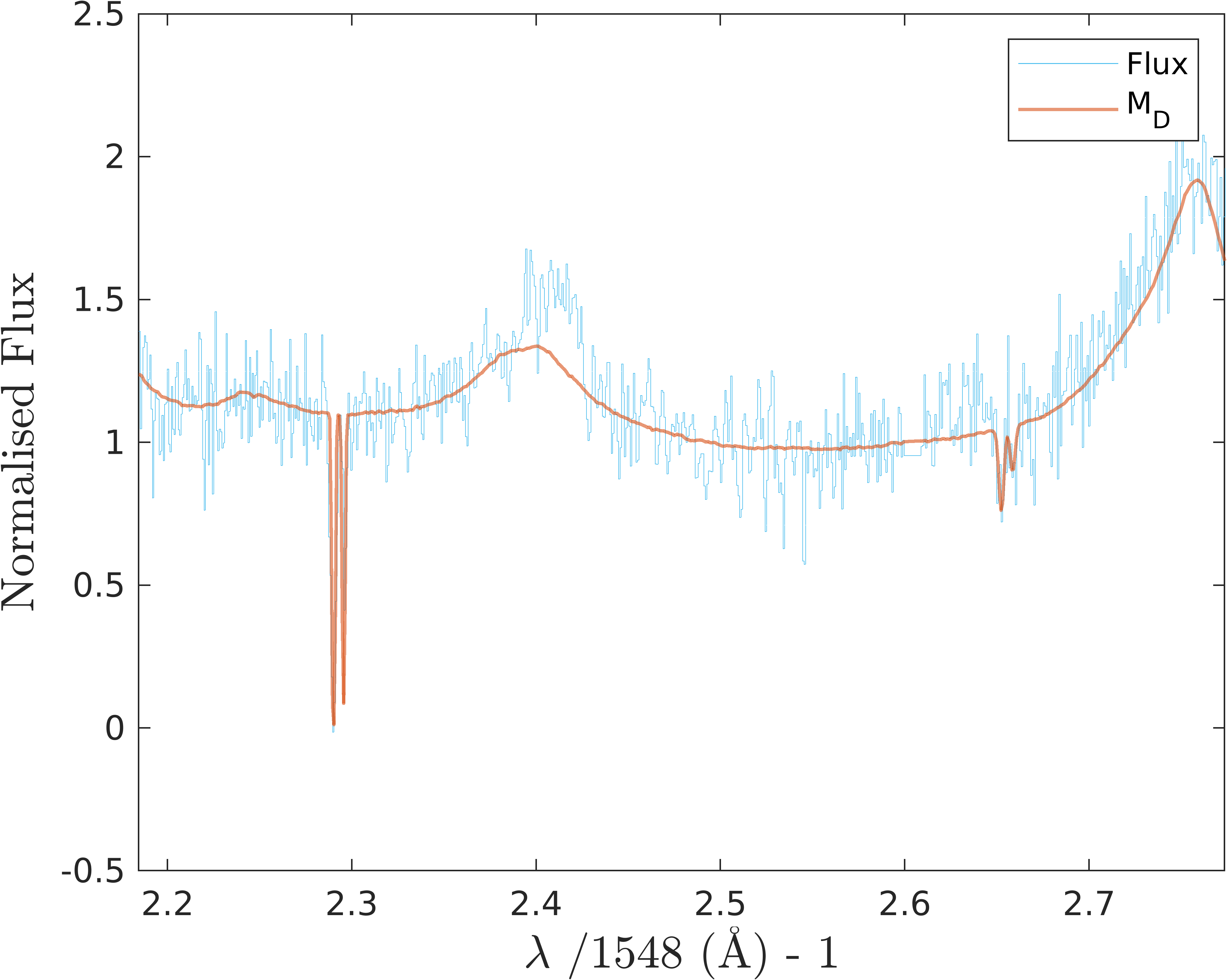}
\caption{Example spectrum with two \civ\ absorbers found by GP with high confidence but not included in the PM catalogue.
The QSO-ID is 51994-0309-592 and $\zqso =  2.76$. Posterior probabilities for the two searches are P$(\model_D)=[1.00, 0.98]$. The maximum \textit{a posteriori} absorption redshifts are $\zciv=[2.288, 2.650]$,
and the rest equivalent widths are W$_{r,1548}^{\rm GP, flux}=[0.90, 0.32]$~\AA. These two ``\civ'' systems are actually non-\civ~absorption lines from a strong, complex system at lower redshift.
The PM pipeline identified the $z=2.288$ lines as a \civ\ \textit{candidate} but ranked it zero; the $z=2.650$ ``\civ\ 1550 \AA'' line
 fell below the PM detection threshold.}
\label{fig:PM0GP1}
\end{figure}

In this section we examine example absorbers from the GP only, GP uncertain and PM only categories discussed above.
Figure \ref{fig:PM0GP1} shows an example of a
spectrum where the GP catalogue shows two absorbers with probability more
than 95\%, but the PM catalogue has zero detections. In this case, the GP \civ\ at $z=2.288$ is
actually \ion{Al}{II} $\lambda1670$ from a strong, multi-component system at $z=2.05$ with
\ion{Mg}{II} $\lambda\lambda2796,2803$; \ion{Fe}{II} $\lambda\lambda\lambda2344,2374,2384$
and $\lambda\lambda2586,2600$; and \ion{Al}{III} $\lambda\lambda 1854, 1862$.
The latter was flagged by the GP algorithm as $\zciv=2.650$.
The $z=2.288$ ``\civ'' was detected
as a candidate in \cite{C13} but visual inspection revealed its true identification; the $z=2.650$ ``\civ'' was
not even a candidate in \cite{C13} because the would-be 1550 line was not
detected by the automated candidate finder (i.e., it fell below their sensitivity threshold).
Thus some of the absorbers in the GP only category are simply missed by the PM pipeline and some are false detections of other doublets.

\begin{figure}
  \includegraphics[width=\linewidth]{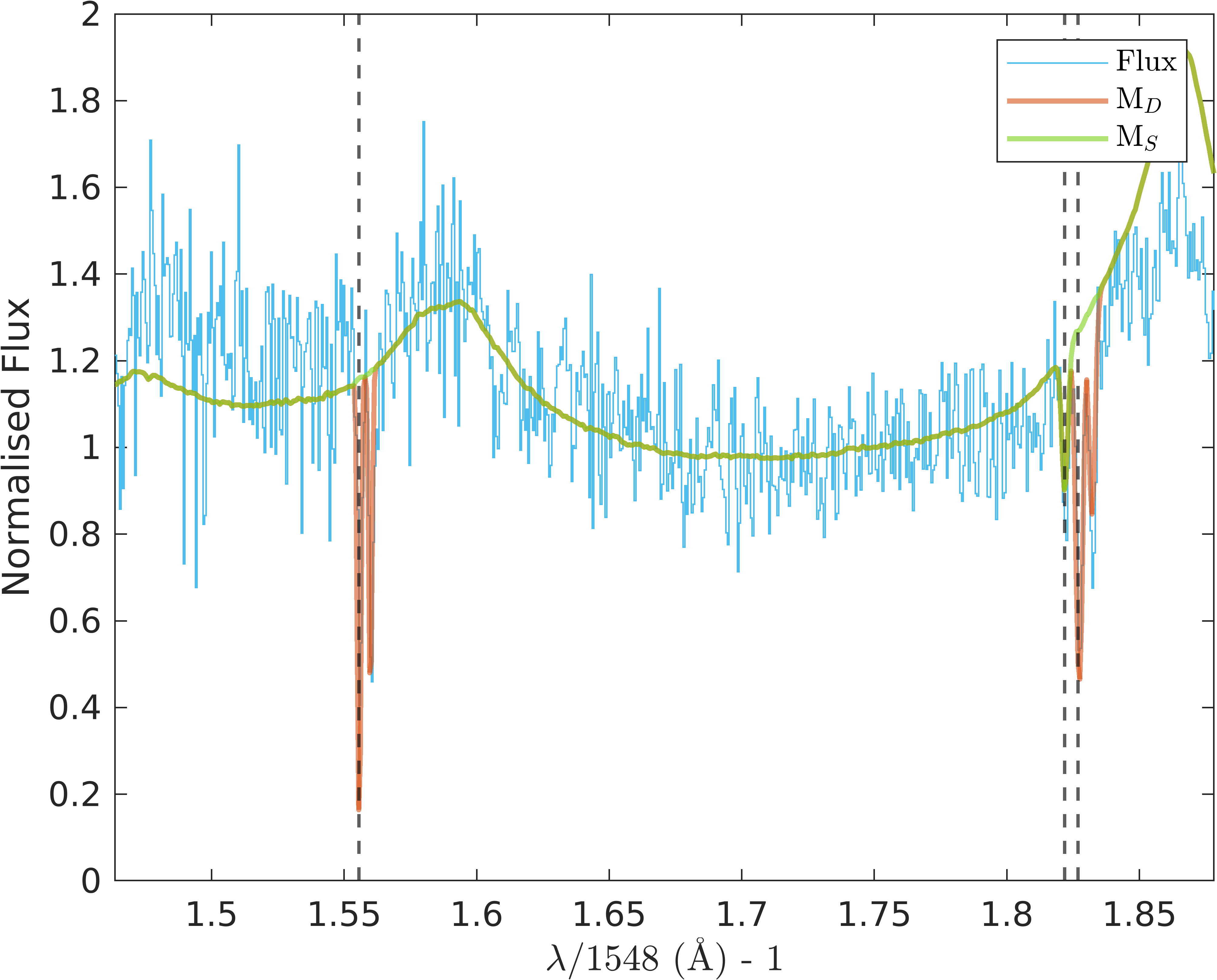} 
  \caption{Example  
  of an absorber at $\zciv^{\rm PM}=1.822$ detected by the PM catalogue, but assigned a relatively low probability (P$(\model_D)=49\%$) by
  the GP catalogue. 
  The QSO-ID for this spectrum is 52367-0332-585, and the quasar redshift is $1.87$. The vertical dashed lines show the position of PM
  absorbers.
  The posterior absorption probabilities are
  P$(\model_D)=[1.00, 1.00, 0.49, 0.15]$, with maximum a posterior absorber redshifts of
  $\zciv=[1.556, 1.827, 1.822, 1.693]$, and the rest equivalent widths are W$_{r,1548}^{\rm GP, flux}=[0.528\pm0.37, 1.21\pm0.25, 0.55\pm0.30, 0.05\pm0.35]~\mbox{\AA}$.
  The PM catalogue reported absorbers at $z_{\rm \civ}^{\rm PM}=[1.556,1.827,1.822]$ with 
  W$_{r,1548}^{\rm PM}=[0.88\pm 0.12, 0.88\pm 0.08, 0.40\pm0.10]$~\AA.}
  \label{fig:PM1GP0}
  \end{figure}

Figure \ref{fig:PM1GP0} shows an example spectrum where the GP pipeline is uncertain about an absorber detected in the PM catalogue. Both pipelines find the \civ~absorber at $\zciv^{\rm GP}=1.827$. However, the PM catalogue identifies a second absorber at $\zciv^{\rm PM}=1.822$. This absorber is  also detected by the GP pipeline. However, the GP pipeline is unable to distinguish between the doublet and singlet models as the $1550$ \AA~line is blended with the higher redshift absorber. It thus assigns both models equal probability, hence P$(\model_D)=49\%$.
%

\begin{figure}
\includegraphics[width=\linewidth]{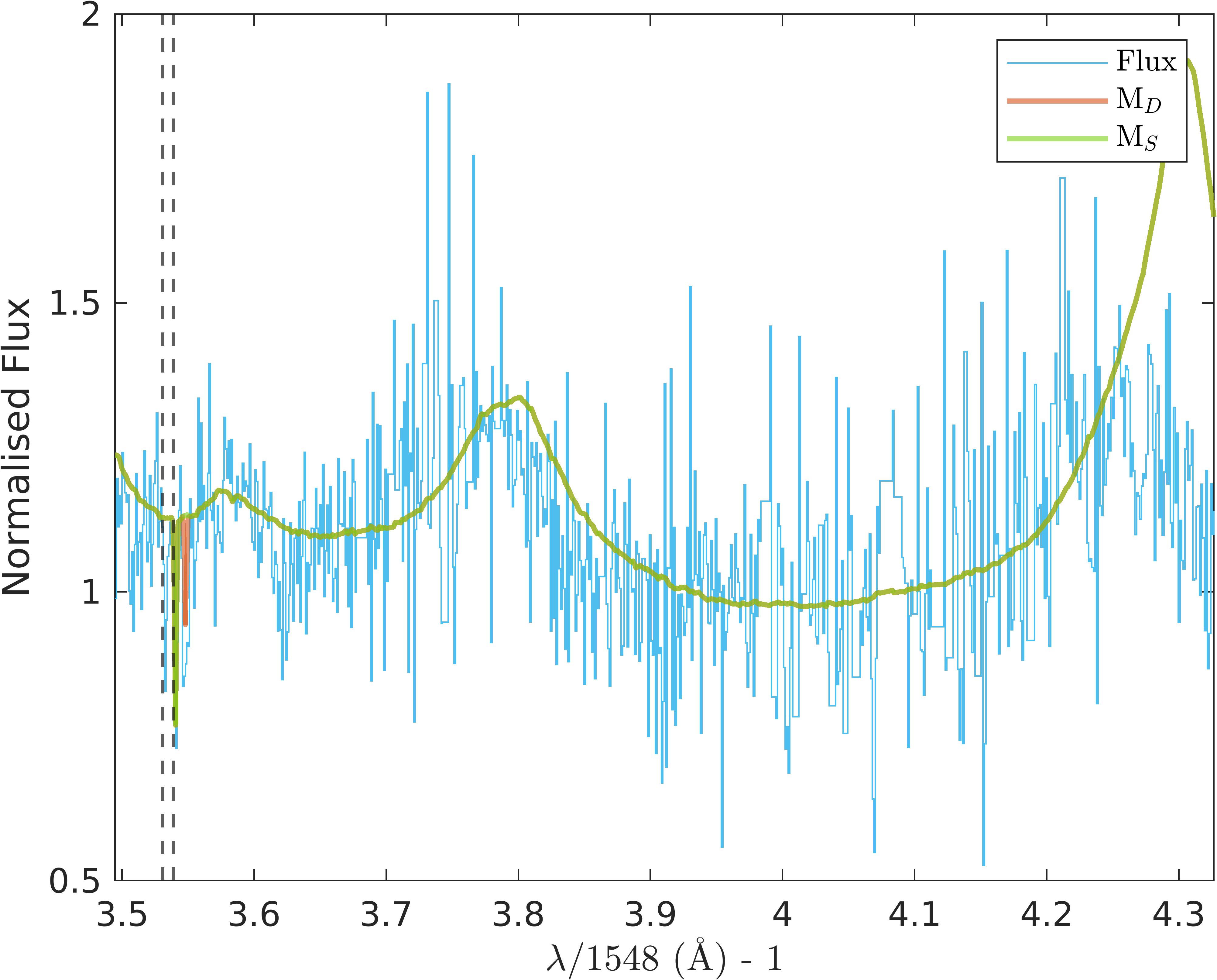}
\caption{Example spectrum containing a PM only absorber for QSO-ID: 51943-0300-475 and
$\zqso=4.31$ where $z_{\rm \civ}^{\rm PM}=[3.5309,3.5389]$ (vertical
dashed lines).  GP assigns P($\model_D)=1$ to
$z_{\rm \civ}^{GP}=3.540574$ which is offset by only $110$ \kms
from $z_{\rm \civ}^{\rm PM}=3.5389$. Before the second search, we mask 350~\kms around the first absorber and thus are unable to detect the second PM catalogue absorber.}
\label{fig:GPOnly}
\end{figure}

Figure \ref{fig:GPOnly} illustrates QSO-ID: 51943-0300-475, which contains an example of an absorber in the PM only category.
The PM catalogue contains two absorbers at
$z_{\rm \civ}^{\rm PM}=[3.5309,3.5389]$. The GP pipeline finds an absorber
in the first \civ-search  at $z_{\rm \civ}^{GP}=3.540574$ with a posterior
probability of 1 for the doublet model. According to our multi-absorber
finding procedure (see Section \ref{sec:multiciv}) we mask the observed flux 350~\kms
around the found absorber and do the next search. However, since the other reported
absorber in the PM catalogue ($z_{\rm \civ}^{\rm PM} = 3.5309$) is offset only 110~\kms
from the absorber found in the first GP search, it is in a masked region and not identified by the GP pipeline in the second search.

\section{Results for SDSS DR12}
\label{sec:results}


We applied our model to find \civ\ absorbers in a subset of the SDSS DR12 quasar catalogue. We searched quasars with rest-frame wavelength coverage between 1310~\AA~and 1548~\AA\
($1.7<\zqso<5.7$), and without detected BALs. This leaves $185,425$ quasar spectra (see Section~\ref{sec:data}).
For each spectrum, the GP pipeline provides (shown as columns in Table \ref{tab:gpTab}):
 posterior probability of \civ\ absorption,
 maximum \textit{a posteriori} values for our absorption model parameters ($\zciv$, $\nciv$, and $\sciv$),
  together with their 95\% confidence intervals, and rest equivalent widths (for 1548~\AA\ and 1550~\AA) and their 95\% confidence intervals.
 Maximum \textit{a posteriori} values and 95\% confidence intervals
 for our absorption model parameters summarise the likelihood distribution,
$P(\Data \vert \theta_i, \zqso, \model_{D})$, of our 10,000 parameter samples (see Equation \ref{eq:model_evidence_qmc}).
Each of these results are contained in a $185,425\times7$ array.
If the search terminated finding fewer than seven absorbers, we report
a \texttt{NaN} value for the columns associated with all further absorbers.
Table \ref{tab:gpTab} shows a snapshot of
our search results for the first 10 absorbers with P$(\model_D)\ge0$.
\begin{table*}
  \caption{For each sight-line, identified by Column 1 and 2, we report the absorber's redshift (Column 3), column density in $\log({\rm cm^{-2}})$ (Column 4),
  Doppler velocity dispersion  in \kms (Column 5), rest equivalent width for 1548 \AA\ W$_{r,1548}$  (Column 6), rest equivalent width for 1550 \AA\ W$_{r,1550}$ (Column 7),  the posterior probability of the \civ\ absorber P($\model_D$) (Column 8), and the posterior probability of the singlet absorber P($\model_S$) (Column 9). We show only absorbers with P($\model_D$)$\neq$\texttt{NaN}.
  This table demonstrates a portion of the full table for the first ten rows. Note that those
  measurements with large errors are uncertain (i.e. low absorption model posterior probability).
  The full table with 445,765 rows is available at \href{https://doi.org/10.5281/zenodo.7872725}{https://doi.org/10.5281/zenodo.7872725}.}
    \begin{tabular}{lllllllll}
    \label{tab:gpTab}
    (1)          &   (2)       & (3)                    & (4)                           & (5)                     & (6)           & (7)           & (8)        &      (9)\\
  QSO-ID         &   $\zqso$   & $\zciv$               & log($\nciv$)                       & $\sciv$                 & W$_{r,1548}$  & W$_{r,1550}$  & P($\model_D$)  & P($\model_S$)\\
                 &             &                       &   $\log({\rm cm}^{-2})$         &  \kms                  & (\AA)          & (\AA)        &  \\     \hline\hline
56238-6173-528   &  2.3091& 1.91039$\pm$0.00037& 15.66$\pm$0.52& 52.24$\pm$0.06& 1.306$\pm$0.960& 1.178$\pm$1.142&0.63& 0.18\\
   &  & 2.21620$\pm$0.00078& 14.75$\pm$0.24& 104.74$\pm$0.13& 1.328$\pm$1.129& 0.838$\pm$0.790& 0.40& 0.00\\
56268-6177-595   &  2.4979& 2.11727$\pm$0.00036& 13.96$\pm$0.22& 58.93$\pm$0.06& 0.280$\pm$0.469& 0.152$\pm$0.294& 0.23& 0.61\\
   &  & 1.99557$\pm$0.00027& 13.99$\pm$0.26& 39.18$\pm$0.04& 0.261$\pm$0.455& 0.148$\pm$0.324& 0.41& 0.00\\
55810-4354-646   &  2.3280& 1.90383$\pm$0.00027& 13.89$\pm$0.11& 48.29$\pm$0.05& 0.230$\pm$0.144& 0.125$\pm$0.083& 0.30& 0.36\\
   &  & 1.94502$\pm$0.00060& 13.77$\pm$0.27& 55.32$\pm$0.10& 0.187$\pm$0.330& 0.099$\pm$0.186& 0.31& 0.00\\
56565-6498-177   &  2.3770& 1.95293$\pm$0.00049& 14.52$\pm$0.34& 49.38$\pm$0.08& 0.636$\pm$0.803& 0.418$\pm$0.678& 0.29& 0.32\\
56268-6177-608   &  3.7120& 3.33339$\pm$0.00033& 14.49$\pm$0.05& 75.29$\pm$0.04& 0.763$\pm$0.174& 0.460$\pm$0.125& 1.00& 0.00\\
   &  & 3.51346$\pm$0.00065& 14.21$\pm$0.17& 96.87$\pm$0.08& 0.530$\pm$0.564& 0.289$\pm$0.345& 0.76& 0.00\\
  &  & 3.22375$\pm$0.00065& 14.06$\pm$0.23& 62.16$\pm$0.08& 0.344$\pm$0.363& 0.189$\pm$0.208& 0.28& 0.00\\

  \end{tabular}
  \end{table*}

\begin{figure*}
  \centering
  \begin{subfigure}[t]{0.47\textwidth}
    \centering
  \includegraphics[width=\linewidth]{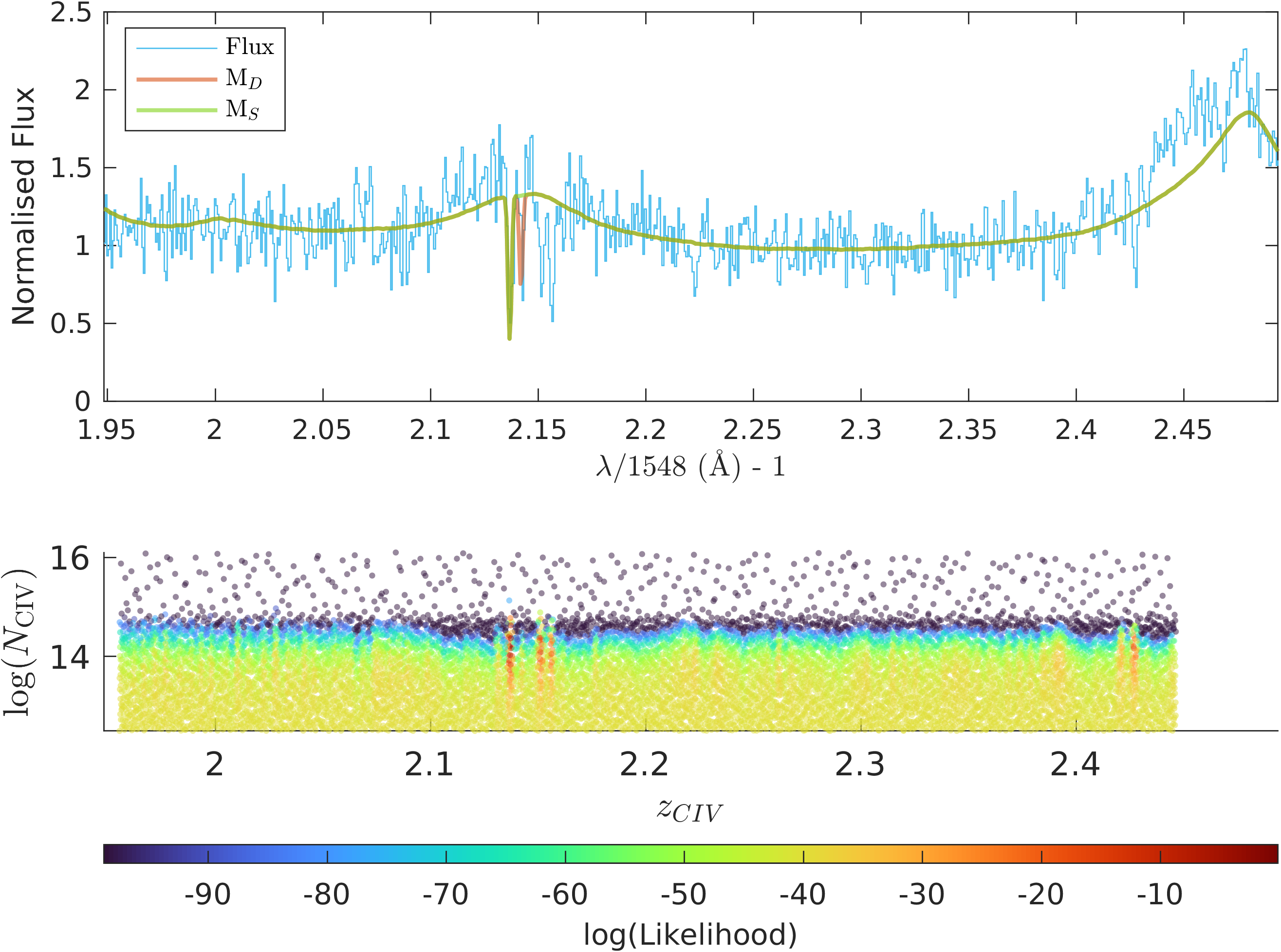}
  \caption{The first \civ~search:
The upper panel shows the normalised flux (light blue),
\civ\ model (M$_D$, red curve), and the single line model (M$_S$, green curve)
as a function of \civ~redshift.
The lower panel shows the likelihood function value for $\model_D$
 as a colour map for each of the 10,000 $\zciv$ samples (x-axis) and
$\nciv$  samples. The third parameter ($\sciv$) is projected onto this 2D space.
Our GP pipeline gives the following results for the first search:
P($\model_D$)=1.00, $\zciv$=2.13682$\pm$0.00049, log($\nciv$)=14.42$\pm$0.20, $\sciv$=64.55$\pm$0.08~\kms,
W$_{r,1548}$=0.568$\pm$0.372 \AA, W$_{r,1550}$=0.072$\pm$0.386 \AA.}
\label{fig:search1}
\end{subfigure}
\hfill
\begin{subfigure}[t]{0.47\textwidth}
\centering
  \includegraphics[width=\linewidth]{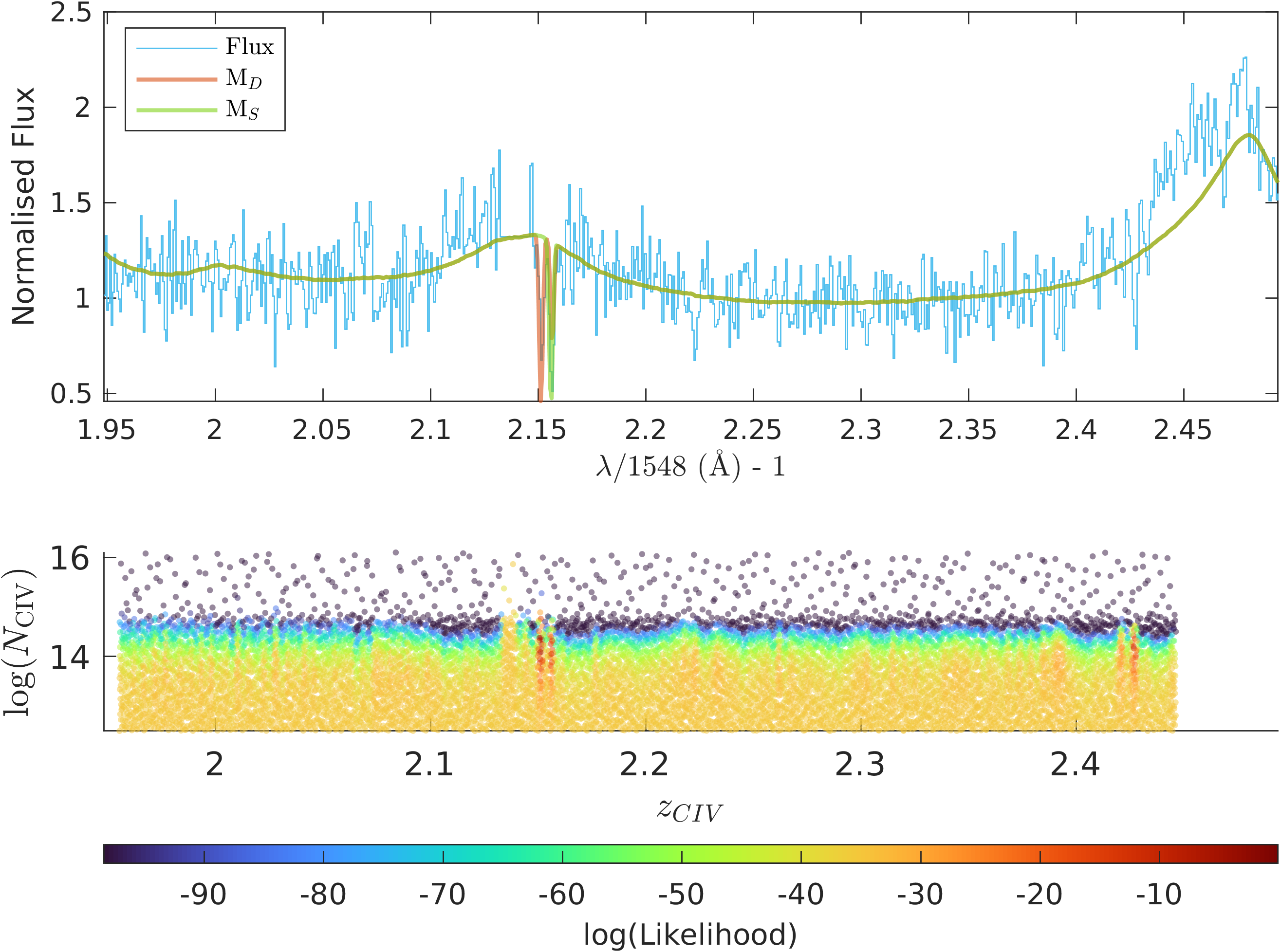}
  \caption{The second  \civ~search:
  The upper panel is similar to Figure \ref{fig:search1}. However,
  we masked 350 \kms around the absorber found in the first \civ~search at $\zciv=2.13682$.
  The lower panel shows the likelihood function values for $\model_D$ after masking the region around the absorber found in the first step.
  Our GP pipeline gives the following results for the second \civ~search:
  P($\model_D$)=1.00, $\zciv$=2.15132$\pm$0.00076, log($\nciv$)=14.38$\pm$0.21, $\sciv$=64.55$\pm$0.08~\kms,
  W$_{r,1548}$=0.615$\pm$0.365 \AA, W$_{r,1550}$=0.707$\pm$0.376 \AA.}
  \label{fig:search2}
  \end{subfigure}
\hfill
\begin{subfigure}[t]{0.47\textwidth}
  \centering
  \includegraphics[width=\linewidth]{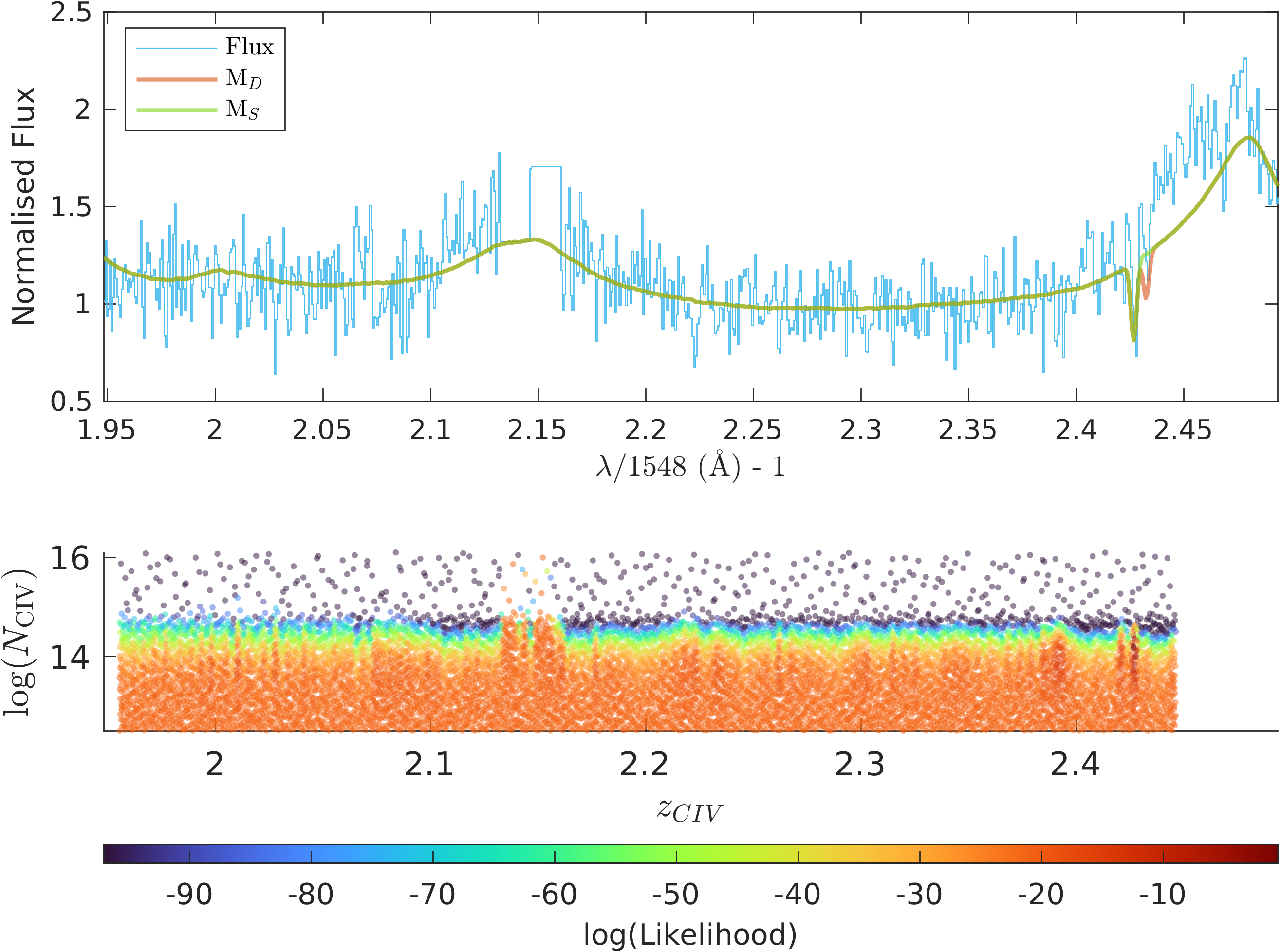}
  \caption{The third \civ~search:
  The upper panel is similar to Figure \ref{fig:search2} but with
  350 \kms around the two absorbers found in the first and second \civ~searches at $\zciv=2.13682$ and $\zciv=2.15132$ masked.
   The lower panel shows the likelihood function value
   as a colour map after masking both absorbers.
   Our GP pipeline gives the following results for the third search:
  P($\model_D$)=1, $\zciv$=2.42670$\pm$0.00006, log($\nciv$)=14.17$\pm$0.02,
   $\sciv$=111.81$\pm$0.01~\kms, W$_{r,1548}$=0.164$\pm$0.407 \AA, W$_{r,1550}$=-0.602$\pm$.396 \AA.}

  \label{fig:search3}
  \end{subfigure}
  \hfill
  \begin{subfigure}[t]{0.47\textwidth}
    \centering
    \includegraphics[width=\linewidth]{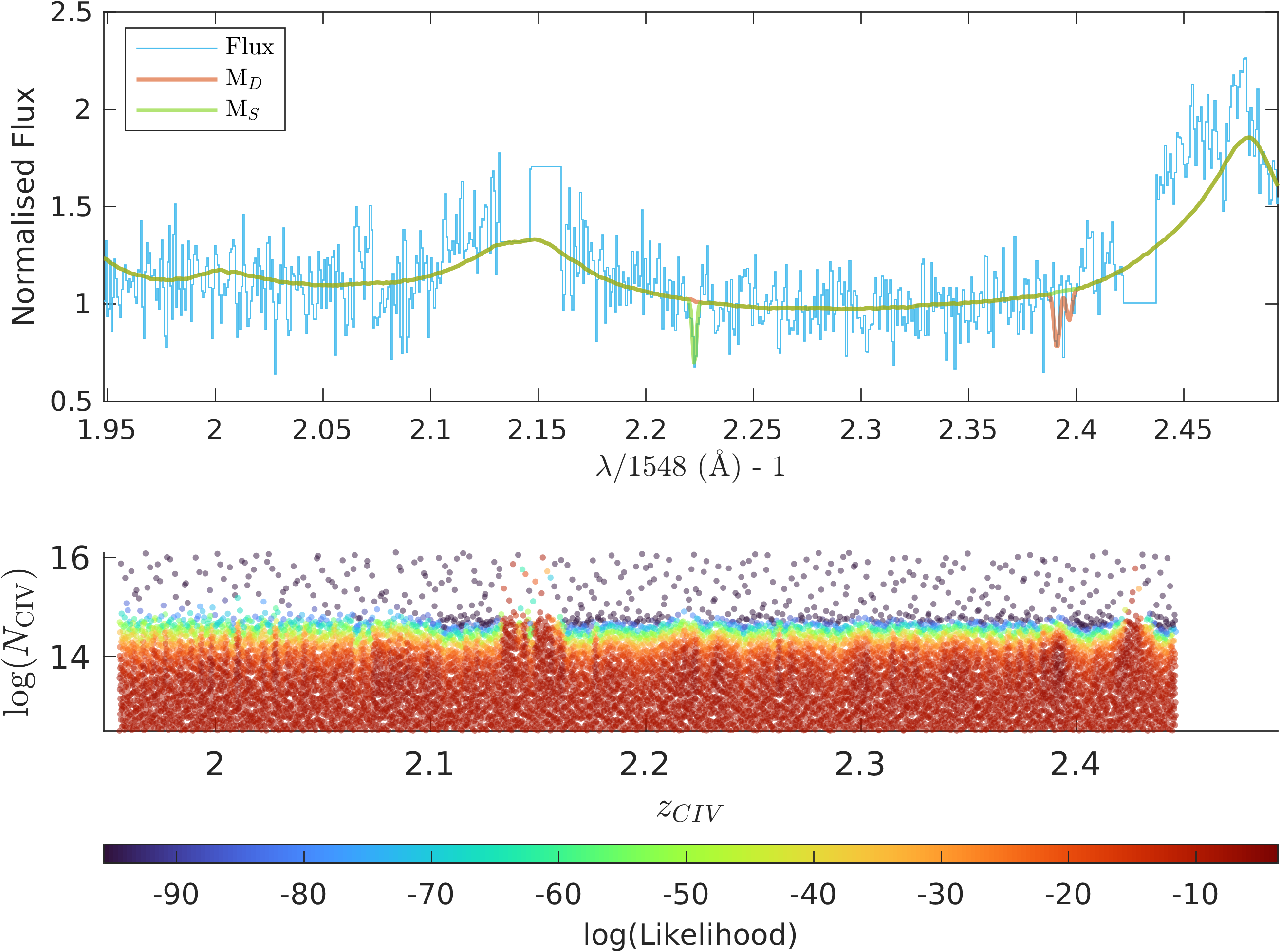}
    \caption{The fourth and final \civ~search.
    The upper panel is similar to Figure \ref{fig:search3} but with 350 \kms around the absorbers found by the previous three searches masked.
    The lower panel shows the likelihood function value
     as a colour map for each of the 10,000 $\zciv$ samples (x-axis) and
    $\nciv$  samples. Our GP pipeline gives the following results for the final search:
    P($\model_D$)=0.27, $\zciv$=2.39100$\pm$ 0.00341, log($\nciv$)=14.04$\pm$0.82, $\sciv$=105.99$\pm$0.56~\kms,
    W$_{r,1548}$=0.248$\pm$0.434, W$_{r,1550}$=-0.085$\pm$0.424.
    Note that since the highest probability in the fourth search was
    P($\model_N$)=0.73, the algorithm performs no further searches (see Section \ref{sec:multiciv}).}
    \label{fig:search4}
    \end{subfigure}
\caption{Panels (a) through (d), show four subsequent searches
for \civ\ absorption on the spectrum of QSO-56265-6151-936 with a redshift of
2.4811.}
  \end{figure*}

Figures \ref{fig:search1} through \ref{fig:search4} illustrate
the four
\civ~searches done by
the GP pipeline on QSO-56265-6151-936 with $\zqso = 2.4811$, and we briefly
explain these iterations here. We found a \civ\ absorber at $\zciv=2.13682$.
In the first search, the null model had $P(\model_{N}) =0.0$, the single
line model $P({\model_S})=0.0$, and the \civ\ doublet model $P({\model_D})=1.0$.
We thus masked the \civ\ doublet model 350 \kms around the \civ~absorber at $\zciv=2.13682$ in the first \civ~search and commenced the second
search, shown in Figure \ref{fig:search2}.
Our second search found an absorber at $\zciv=2.15132$ with $P({\model_D})=1.0$,
$P({\model_S})=0.0$, and $P({\model_{N}})=0.0$.
For the  third \civ~search we masked 350~\kms around each of the absorbers found in the previous
steps and found a third absorber at $\zciv=2.42670$, again with
with $P({\model_D})=1.0$, $P({\model_S})=0.0$, and $P({\model_{N}})=0.0$. The fourth search, with regions around all three previous absorbers masked, found P$({\model_D})=0.27$,
P($\model_S)= 0.0$ and P$(\model_N)=0.73$. Since the null model now had the largest model posterior, this was the final \civ~absorber search in this spectrum (see Section \ref{sec:multiciv}).

\begin{figure}
  \includegraphics[width=\linewidth]{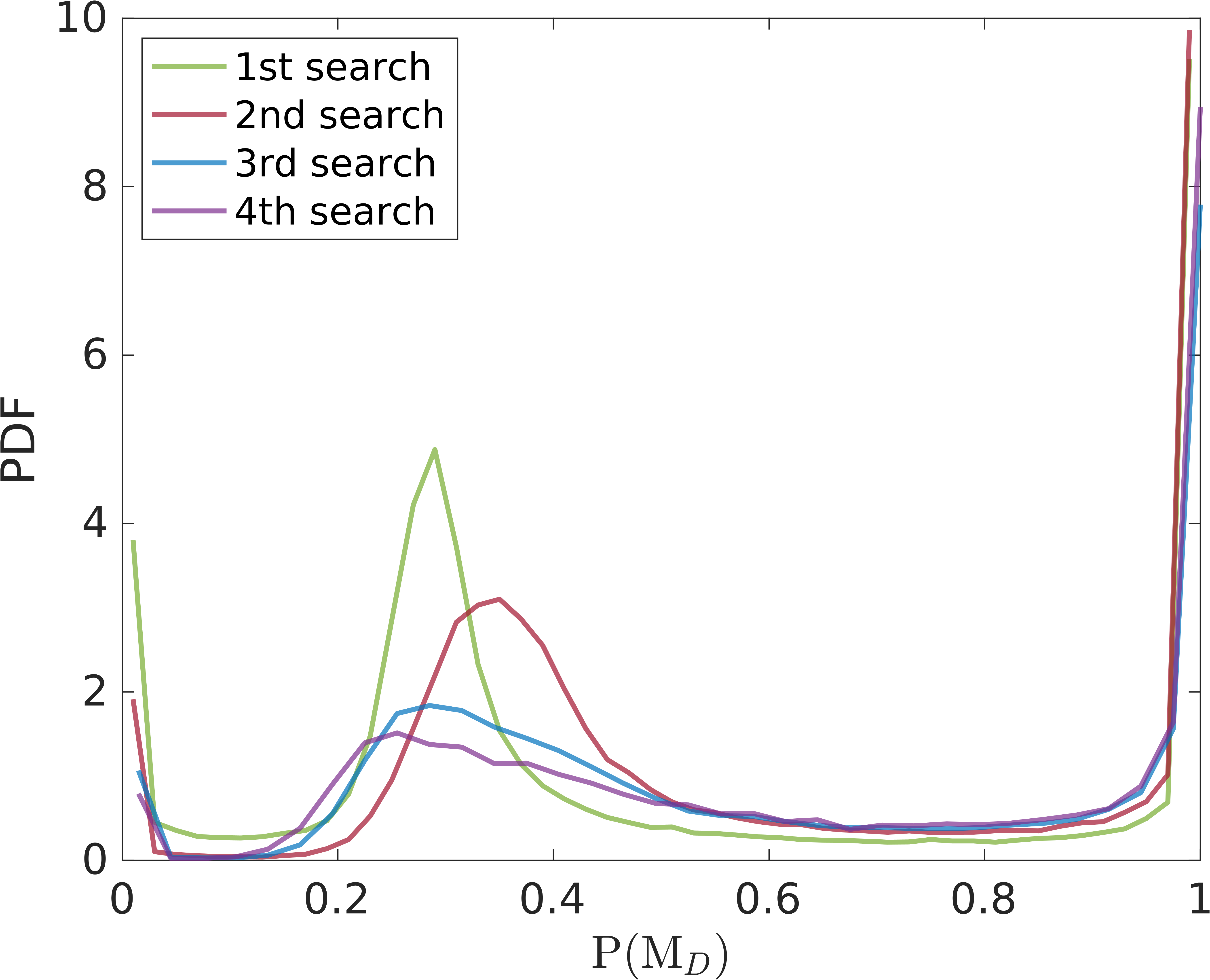}
  \caption{Distribution of P$(\model_D)$ for the first to fourth searches.
  We do not show the fifth to seventh searches as they find very few absorbers (see Table \ref{tab:p_c4}). The peak around P($\model_D)\sim 0.3$ comes from low SNR spectra where the posterior probabilities of our three models are dominated by their priors.}
  \label{fig:distp1}
\end{figure}

Looking at the distribution of each  model posterior probability
 gives us an insight into
how the spectra have been classified.
Figure~\ref{fig:distp1} shows the distribution of doublet model posterior probabilities for the first four \civ\ absorber searches in the DR12 spectra.
For each search, we see a peak in the posterior probability
distribution around 30\%, stronger for earlier searches.
We also examined ${\rm P}(\model_S)$ and
${\rm P}(\model_{N})$ for the first search and found that many of these ambiguous \civ~absorbers
also have ${\rm P}(\model_S) \sim 0.3$. In addition, these
absorbers are generally in lower S/N spectra. Thus this peak occurs when the spectra are weakly constraining and the posterior absorber probability is dominated by the model priors.

Table \ref{tab:p_c4} summarises the reported posterior probabilities for our catalogue.
Around 66\% of spectra have no \civ\ absorbers detectable at more than 85\% confidence.
Around 15\% of spectra have one doublet and around 8\% two doublets, each with a confidence more than 85\%.
The probability for detecting two independent absorbers in a spectrum is:
\begin{equation}
 {\rm P}(2~\civ)= {\rm P}(1~\civ) \times {\rm P}(1~\civ) = 0.15^2 \sim 2.2\%.
\end{equation}
The actual probability of two \civ\ absorbers in a spectrum is higher, $\sim$ 8\%, demonstrating that absorbers are not independent but strongly correlated. Furthermore, we find five or more doublets at $> 85\%$ confidence in 3.1\% of spectra.

We detected a single line absorber in $\sim 10$\% of sight-lines.
If single line absorbers were independent, we
would expect $0.1^2$ or 1\% of spectra to contain two singlet line absorbers. The actual probability of finding two singlets in a single sight-line was $\sim 2\%$, so the correlation between single line absorbers is much weaker than for \civ.

%
\begin{table}  
  \caption{The number of spectra containing different numbers of \civ~absorbers for various doublet model probability thresholds, P($\model_D$).
  The first column shows the number of \civ\ absorbers found within each spectrum (see Section \ref{sec:multiciv}) .
   The second through fourth
   columns show probability thresholds of > 65\%, 85\%, and 95\% respectively. Cells show the number of quasar spectra falling in each category, together with
   the corresponding percentage of the 185,425 spectra in our SDSS DR12 sample.
   } 
  \label{tab:p_c4}
    \begin{tabular}{|l|l|l|l|}
    \hline
    N(\civ) &	P$(\model_D)$>0.65 &	P$(\model_D)$>0.85 &	P$(\model_D)$>0.95\\ \hline\hline
    0 &	113142 &	123994 &	131767 \\
     &  61.0\% &  66.8\% & 71.0\% \\ \hline
    1	& 31163	& 27733	 & 24981 \\
      & 16.8\% & 14.9\% & 13.5\% \\\hline
    2	& 17526 &	14533 	& 12777 \\
     & 9.4\%  & 7.8\% & 6.9\% \\ \hline
    3	& 10020	& 8424 &	7218 \\
    	& 5.4\% & 4.5\% &	3.9\%\\ \hline
    4&	6112 &	4960 &	4176 \\
    &	3.3\%&	2.6\%&	2.3\%\\ \hline
    5	& 4155 	& 3342	& 2656 \\
    	& 2.2\%	& 1.8\%& 1.4\%\\ \hline
    6&	2426	 & 1771 &	1348 \\
    &	1.3\% & 0.9\% &	0.7\%\\ \hline
    7& 881 & 668  & 502 \\
    &  0.5\%&  0.4\% & 0.3\%\\
      \end{tabular}
\end{table}

\begin{figure}
  \includegraphics[width=\linewidth]{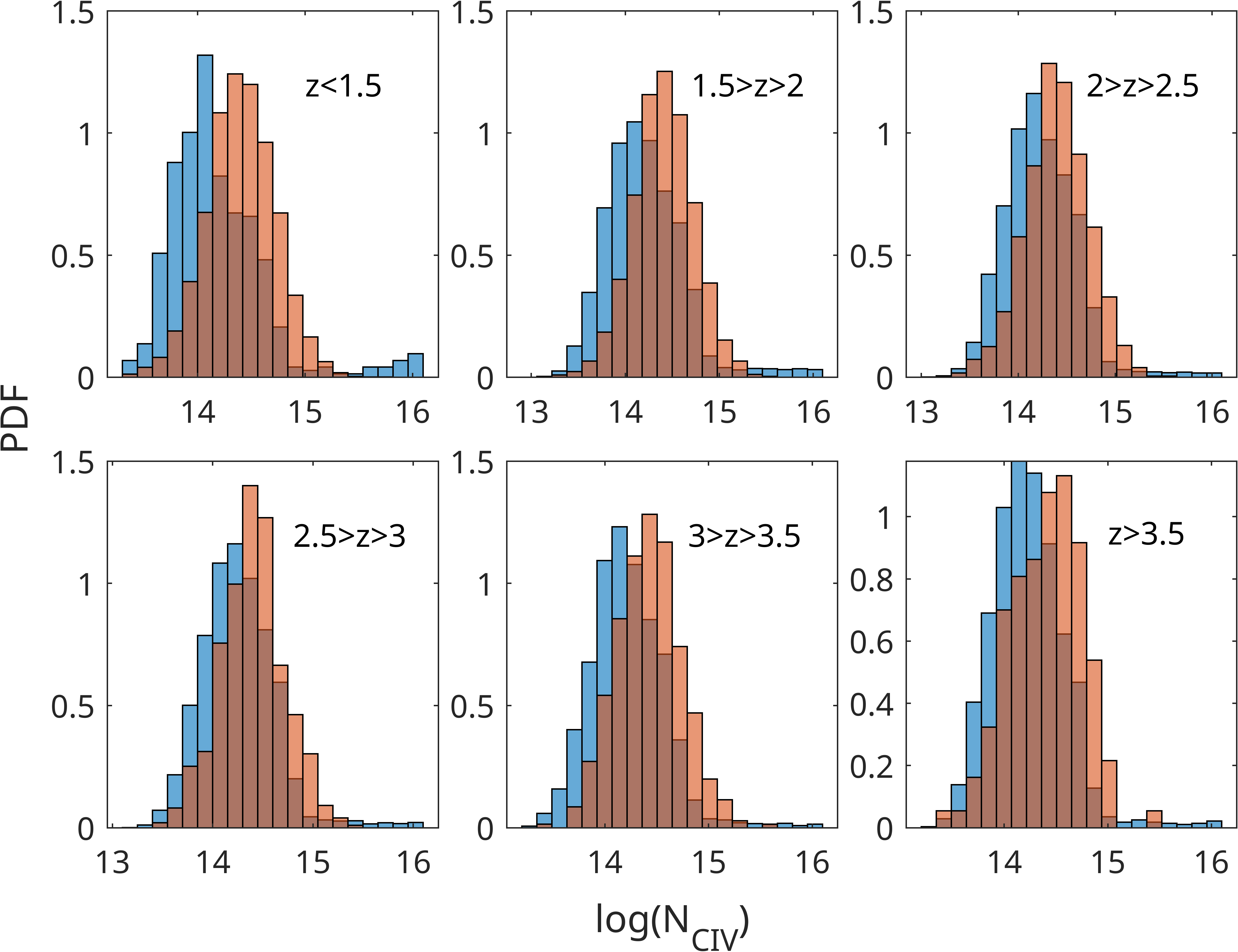}
  \caption{Column density statistics for our GP results (blue histograms) compared to
the training \civ\ catalog from SDSS DR7 (red histograms). Each panel is showing a 
specific redshift bin. Y-axes show the noralized probability distribusion function (PDF).
Please note that column density values should be considered as lower limits (\citep{C13})
because they lines are partially to completely saturated so we can only measure lower limits.}
\label{fig:colDenCompar}
\end{figure}

We provided the distribution of maximum a posteriori column densities 
for with P($\model_D)>0.95$ in Figure \ref{fig:colDenCompar} for 
different redshift bins.
Blue histograms in each panel correspond to the probability
distribustion function (PDF) from the GP pipeline while the red histograms
show the column densities in the PM catalogue. Note 
that column density values measured by our pipeline are often lower limits because 
given the low resolution spectra of SDSS, they are partially to completely saturated. 

\civ\ absorbers have a power law distribution \citep{Ellison2000}, so that we
should observe more absorbers for lower equivalent widths. However, weaker absorbers are harder to detect due
to observational limitations. That is why the distribution of 
column densities in all of the panels of Figure \ref{fig:colDenCompar}
starts to drop for $\log(\nciv) \lesssim 14$.
This turning point allows a rough estimate for the
completeness limit of our catalogue for detecting weak absorbers.
The turning point is at lower column densities for the blue histograms,
indicating that the GP pipeline is slightly more sensitive than the
PM catalogue.

One can do a completeness test for the detection of 
column density or rest equivalent width  by randomly injecting 
synthetic absorbers into absorption-free quasar spectrum and
assessing the probability of recovering those artificial
absorbers \citep{C13}. 
Panels of Figure \ref{fig:colDenCompar} shows again that the completeness
for column density is around $\log(\nciv) \sim 14$. In a follow up project
we will perform a detailed analysis of the completeness of our technique, including
any possible redshift or signal to noise dependence.

\begin{figure}
\includegraphics[width=\linewidth]{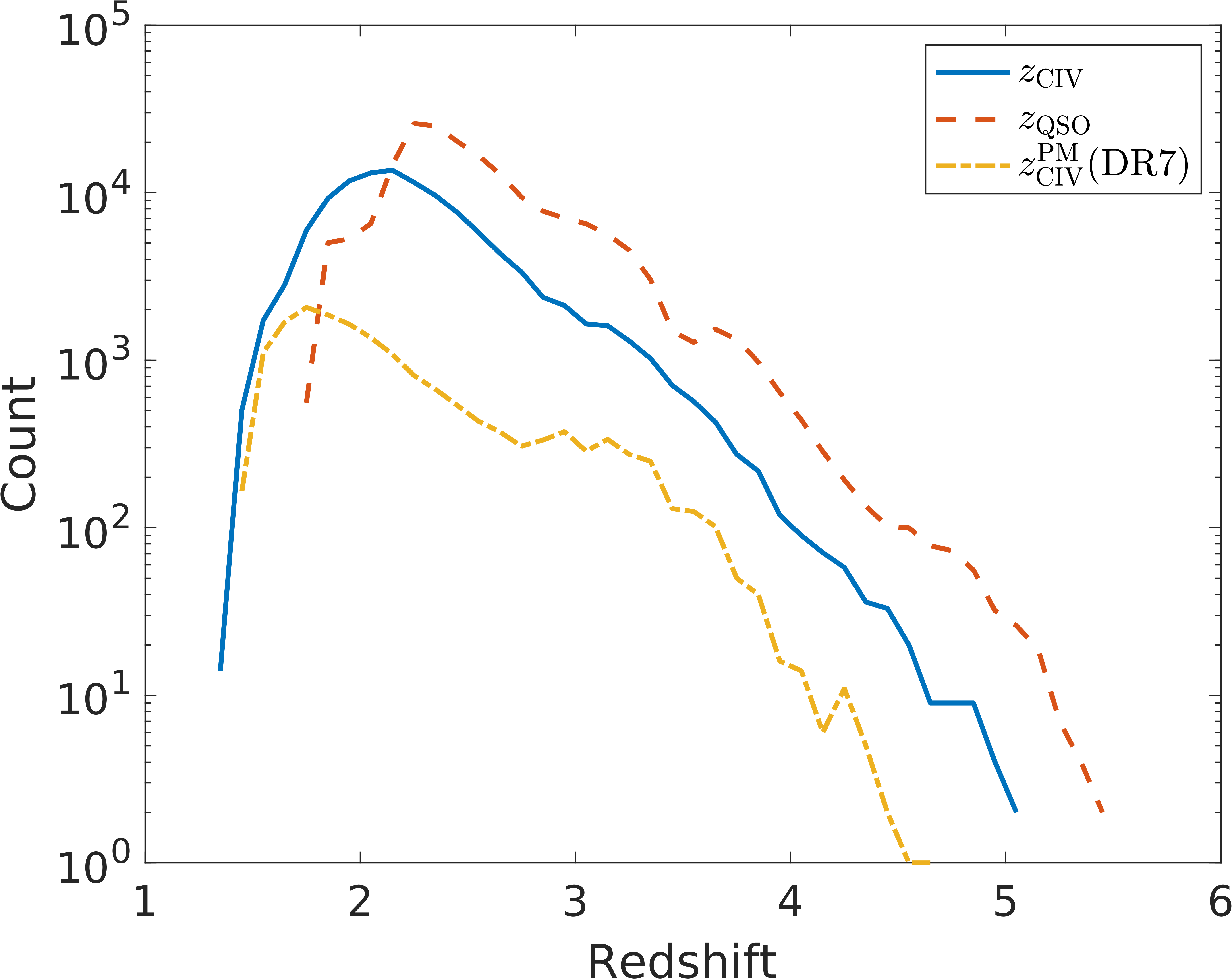}
\caption{The redshift distributions of DR12 quasars (red dashed), high-probability (P($\model_D$)$\ge0.95$) DR12 GP \civ\ absorbers
 (blue solid), and DR7 PM \civ\ absorbers with ranking$\ge$2 (yellow dot-dashed). The quasar redshift is offset towards redder values than the absorber redshift, as expected, since absorbers cannot be more redshifted than the quasar. The GP catalogue finds absorbers outside of the absorber redshift range reported in the PM catalogue.
}
\label{fig:DR12Z}
\end{figure}

Figure \ref{fig:DR12Z} shows the distribution of (maximum \textit{a posteriori}) absorber redshifts.
There is a peak around $z\sim 2$, mirroring the distribution of quasar
redshifts. Overall, we have detected an order of magnitude more absorbers
than the PM catalogue, reflecting the larger size of our sight-line sample.
There are $33$ absorbers in DR12 with  ${\rm P}(\model_D)>0.85$
and a redshift higher than $4.68$, the maximum reported
$\zciv$ in the PM catalogue.

\begin{figure}
\includegraphics[width=\linewidth]{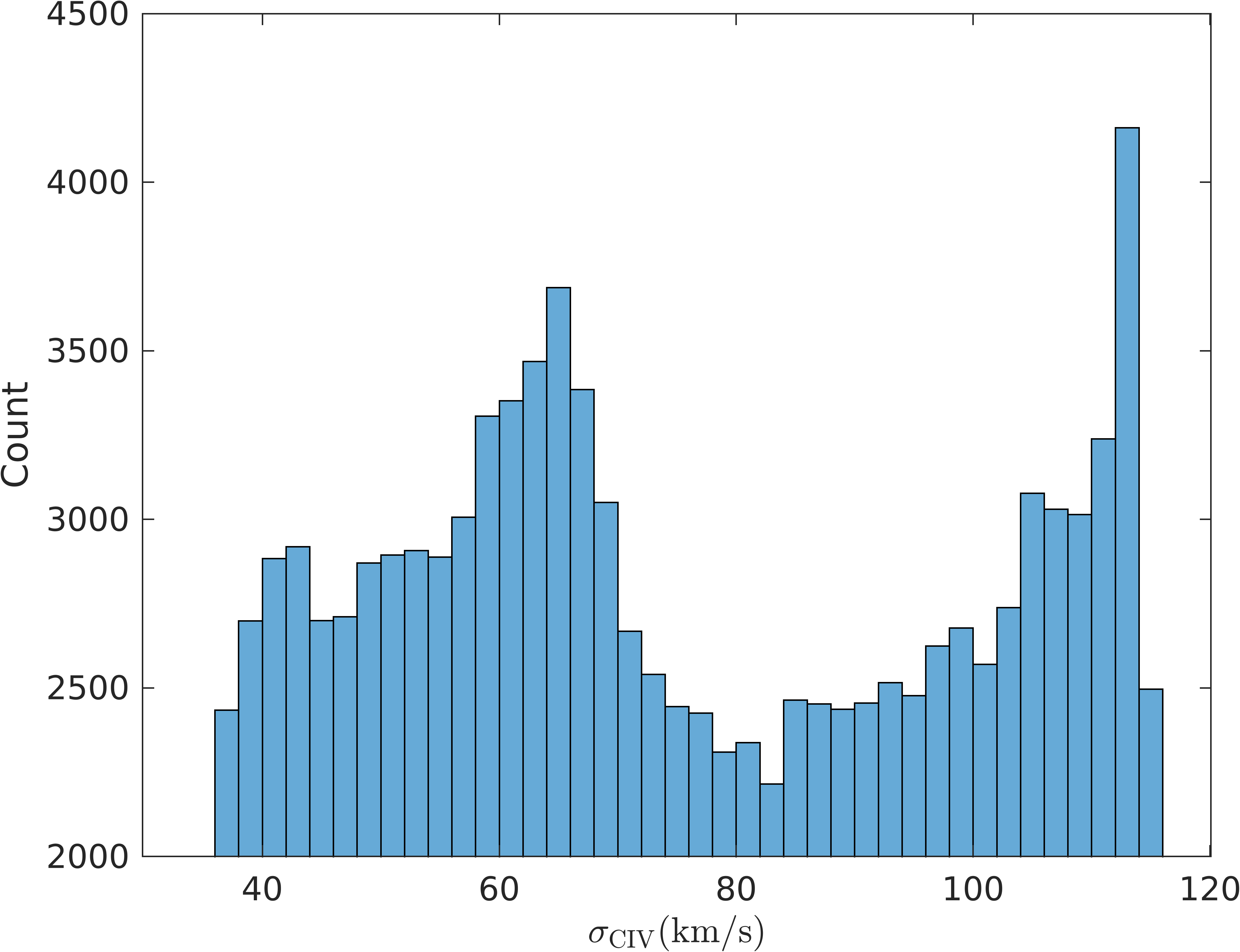}
\caption{Distribution of the maximum \textit{a posteriori} Doppler velocity dispersion values for absorbers detected in SDSS DR12 with P$(\model_D)\ge0.95$. Our prior distribution for Doppler velocity
dispersion was uniform between 35~\kms and 115~\kms but the posterior
distribution is bimodal. The larger $\sciv$ posterior values are mostly associated with \civ~absorbers found near low SNR pixels.}
\label{fig:DR12Sigma}
\end{figure}

In Figure \ref{fig:DR12Sigma} we show the distribution
of maximum \textit{a posteriori}
Doppler velocity dispersion, $\sciv$, for the absorbers detected in the DR12 spectra. While
the adopted prior for $\sciv$ was a flat distribution between 35~\kms and
115~\kms, we see that the posterior distribution is moderately bimodal.
The peak at larger $\sciv$ values is connected with
larger column densities. We examined the spectra of detected
absorbers with $\log\nciv > 16$ and $\sciv > 110$~\kms. Most of these spectra are noisy and in some
 cases the line is heavily blended.
The mean S/N in the region of \civ\ search for absorbers 
with probability larger than 
85\% is $6.4~{\rm pix}^{-1}$, 
compared to a mean S/N of $5~{\rm pix}^{-1}$ for the search region 
of \civ\ absorbers in the spectra
that contain absorbers with $\sciv>80$ \kms and $\log\nciv>15$.

\begin{figure}
  \includegraphics[width=\linewidth]{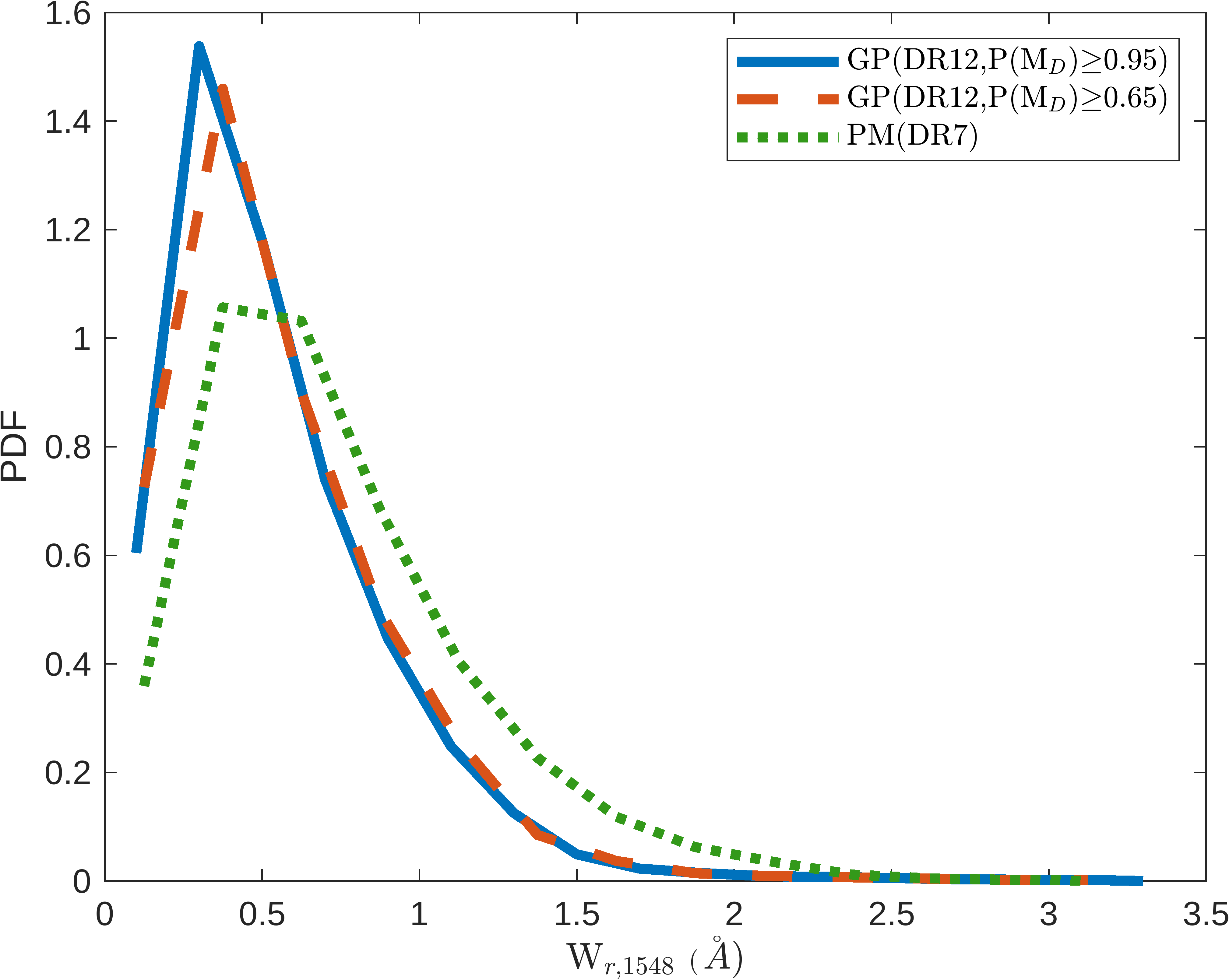}
  \caption{The distribution of the rest equivalent width of the 1548~\AA\ (${\rm W}_{r,1548}^{\rm GP, Voigt}$) line obtained by Voigt profile
  integration (Equation \ref{eq:rewgp}).
  We show ${\rm W}_{r,1548}^{\rm GP, Voigt}$ for all detected
  DR12 absorbers with  P$(\model_D)\ge0.95$ (blue curve) and P$(\model_D)\ge 0.65$ (dashed red curve).
  We also show for comparison ${\rm W}_{r,1548}^{\rm PM}$ values from the PM (DR7) catalogue in the
  dotted green curve.}
  \label{fig:DR12-DR7-REW}
  \end{figure}

Figure \ref{fig:DR12-DR7-REW} shows the 1548 \AA\ rest equivalent widths from our
SDSS DR12 catalogue. We use rest equivalent widths derived from the parameters of the Voigt profile doublet model, $W_{r,1548}^{\rm GP, Voigt}$, which are computed using:
\begin{equation}
W_{r,1548}^{\rm GP, Voigt} =  \int (1 - a_{1548})d\lambda. 
\label{eq:rewgp} 
\end{equation}
Here $a_{1548}$ is the absorption function for our 1548~\AA\ line,
and we compute the rest equivalent width from the maximum \textit{a posteriori} values of $\zciv$, $\nciv$, and $\sciv$ under $\model_{D}$.
Figure \ref{fig:DR12-DR7-REW} also shows the DR7 PM catalogue rest equivalent width distribution for comparison.
The larger sample of spectra in SDSS DR12 allows us to probe higher rest equivalent widths, with 110
absorbers at larger rest equivalent widths than 3.15~\AA, the highest  rest equivalent width in the PM catalogue.
The 50\% completeness limit for ${\rm W}_{r,1548}$ in the PM catalogue is 0.6~\AA~\citep{C13}. Figure \ref{fig:DR12-DR7-REW} suggests
that our catalogue is reasonably complete for ${\rm W}_{r,1548} > 0.4$ \AA. We will perform a detailed statistical analysis, including
the completeness for rest equivalent widths, in a follow-up paper.



\section{Conclusions}
\label{sec:conclusion}
We trained a quasar continuum model to detect \civ~absorbers using a Gaussian process. The training was done on a sample of DR7 spectra which were labelled as \civ~free in the Precious Metals catalogue of \cite{C13}. We used Bayesian model selection to compare our continuum model to models containing one to seven \civ~doublets. 
We added an extra model for single line absorbers, to avoid confusion from interloping metal lines.
The prior distribution was taken from our training catalogue and flat parameter priors were used for the \civ~redshift, $\zciv$ and Doppler velocity dispersion $\sciv$. We searched for up to 7 absorbers in each tested spectrum and provide a comprehensive catalogue containing \civ~detection probability, as well as maximum a posteriori values and credible intervals for $\zqso$, $\nciv$ and $\sciv$. We validated our pipeline by applying it to a hold-out sample of $1301$ spectra from the PM catalogue. Our pipeline produced similar results to the PM catalogue and has good purity and completeness. Generally the two catalogues produced similar \civ~redshifts and rest equivalent widths.

Thus validated, we applied our model to SDSS DR12, and produced the largest \civ\ absorption catalogue yet seen.
Among the total 185,425 selected quasar spectra in SDSS DR12, we found 113,775
\civ\ doublets with > 95\% confidence. Note that the user may pick the desired confidence threshold in our catalogue, thanks to our reported posterior probabilities for each absorber.
We detected \civ~absorption up to $z \sim 5$, including $33$
 systems at higher redshift than seen in DR7. We also detect 110 absorbers
 in DR12 with a rest equivalent width larger than the maximum in the DR7 catalogue.

This paper presents the first machine learning approch  for detecting \civ\ absorption
lines in quasar spectra. Our GP pipeline models a quasar spectrum by learning the covariance between
different parts of quasar spectra, using an absorber-free training set. Then analytic
Voigt profiles are used to find the absorbers. Other machine learning methods, such as convolutional neural networks,
focus on modeling the absorbers without modeling the quasar continuum (eg. \cite{dlaCNN}).
Moreover, methods based on Convolutional Neural Networks have a considerable
number of hyperparamters, scaling with the number of layers in the network. While Convolutional Neural Networks
are very promising, Gaussian Processes are easier to interpret and come with a natural estimate of uncertainty.
GP pipeline proved that is able to detect Damped Lyman-$\alpha$ absorbers in a Bayesian manner
by considring quasar redshift uncertainty which is usually considered as an  exact parameter \citep{fauber2020}.
This represents a distinct advantage of
the GP model, as current neural network-based absorber finders do not possess a
straightforward approach for marginalizing QSO redshift uncertainty.

Our method is good for detecting unblended \civ\ absorbers. 
Note however, that our GP pipeline contains models only for singlet and doublet absorbers, and so may mis-classify blended or mini-BAL systems, which do not strongly resemble the models it uses.
In these rare cases, when the absorption systems are blended together, the line may be a poor match to both the singlet and doublet models.
In these cases, our pipeline is sometimes unable to distinguish between genuine \civ~doublet absorption and other interloper metal lines.
It is possible that two 
single lines, with the same rest-frame wavelength separation as \civ\, coexist in a spectrum, causing our algorithm to
 misidentify them as a \civ\ doublet.
An example for this case is \ion{O}{I} 1302\AA\ and \ion{Si}{II} 1304\AA\ pair. To avoid 
this specific example, we start our \civ\ search redward of 1310\AA.
Our algorithm could potentially be confused by other similar situations still in our 
wavelength search region. However, since our model uses
a doublet Voigt profile designated for \civ\ absorption,
it does not easily mistake doublet metal lines with other wavelength separations as a \civ\ doublet, such as
\ion{Si}{IV} $\lambda\lambda$ 1393~\AA, 1402~\AA. 

We could have considered a mixture of various
absorbers and then properly marginalise different types of absorbers in the Bayesian model
selection. However, our primary objective in this paper is to efficiently 
identify \civ\  candidates for a large number of spectra. Introducing this level of complexity
into the search pipeline would significantly slow down the process.
Nevertheless, we have provided the catalog containing all posterior samples and all the
files to reproduce the inference. This allows interested users the exibility to pinpoint
specific \civ\  absorber candidates of interest and subsequently recalculate the uncertainty
associated with other types of metal lines as needed. An advantage of our pipeline is its
modularity, enabling users to seamlessly incorporate new absorbers into the code by
adding the corresponding Voigt profile.

Here we used a single Voigt profile to fit the absorption systems. 
The Doppler velocity dispersion prior distribution for our absorption models
ranges from 35 km/s to 115 km/s, which translates to a temperature range of
 $\gtrsim 10^5$ K to $\gtrsim 10^6$ K. 
We use a wide prior on the IGM/CGM temperature to:
1) Be able to represent blended/complex absorption systems that could have been modeled by a combination of Voigt profiles as 
a single Voigt Profile. 2) Keep our method simple. In the future work, we will modify the algorithm so 
that it can better model blended systems and so reliably estimate the temperature of the absorbing environment.

Potential applications of our catalogue include: 1) finding targets for high-resolution follow-up of complex \civ~systems \citep[e.g.~][]{muse2023} 2) Cross-matching with galaxy catalogues to find the properties of the galactic circumgalactic medium within which our \civ\ absorbers lie.
3) Cross-matching with a Damped Lyman-$\alpha$ catalogue to investigate the relationship between the highly ionised carbon and neutral hydrogen in the circumgalactic medium.

Finally, the statistical properties of our
catalogue can be computed and compared to the outputs
of cosmological simulations to test and improve models for galactic feedback.

Our technique can also be applied to later, larger quasar catalogues such as those from the SDSS DR16 and the upcoming Dark Energy Spectroscopic Instrument quasar survey.


\section*{Acknowledgement}

R.M. was supported by Higher Education Emergency Relief Funds. 
RM thanks Fred
Hamann for supporting him for part of this work from NSF grant
AST-1911066. R.M. also appreciates Farhanul Hasan's time for his 
discussions about various topics throughout this project. 
S.B. was supported by NASA ATP 80NSSC22K1897.
We used the HPCC cluster at UC Riverside and AWS credits provided under an amazon machine learning research award. M.H. was supported by NASA FINESST grant 80NSSC21K1840. KLC acknowledges partial support from NSF AST-1615296.
Funding for the SDSS and SDSS-II has been provided by the Alfred
P. Sloan Foundation, the Participating Institutions, the National
Science Foundation, the U.S. Department of Energy, the National
Aeronautics and Space Administration, the Japanese Monbukagakusho, the
Max Planck Society, and the Higher Education Funding Council for
England. The SDSS Web Site is \url{http://www.sdss.org/}.

The SDSS is managed by the Astrophysical Research Consortium for the
Participating Institutions. The Participating Institutions are the
American Museum of Natural History, Astrophysical Institute Potsdam,
University of Basel, University of Cambridge, Case Western Reserve
University, University of Chicago, Drexel University, Fermilab, the
Institute for Advanced Study, the Japan Participation Group, Johns
Hopkins University, the Joint Institute for Nuclear Astrophysics, the
Kavli Institute for Particle Astrophysics and Cosmology, the Korean
Scientist Group, the Chinese Academy of Sciences (LAMOST), Los Alamos
National Laboratory, the Max-Planck-Institute for Astronomy (MPIA),
the Max-Planck-Institute for Astrophysics (MPA), New Mexico State
University, Ohio State University, University of Pittsburgh,
University of Portsmouth, Princeton University, the United States
Naval Observatory, and the University of Washington.

\section*{Data availability}

All of our codes are available
publicly in  \href{https://github.com/rezamonadi/GaussianProcessCIV}
{\texttt{GitHub}}
\footnote{\href{https://github.com/rezamonadi/GaussianProcessCIV}{https://github.com/rezamonadi/GaussianProcessCIV}}
and our final catalogue
can be found in  \href{https://doi.org/10.5281/zenodo.7872725}{\texttt{Zenodo}}.
\footnote{\href{https://doi.org/10.5281/zenodo.7872725}{https://doi.org/10.5281/zenodo.7872725}}

\bsp	
\label{lastpage}

\appendix
\section{Rest Equivalent Width Estimates}

\begin{figure}
  \includegraphics[width=\linewidth]{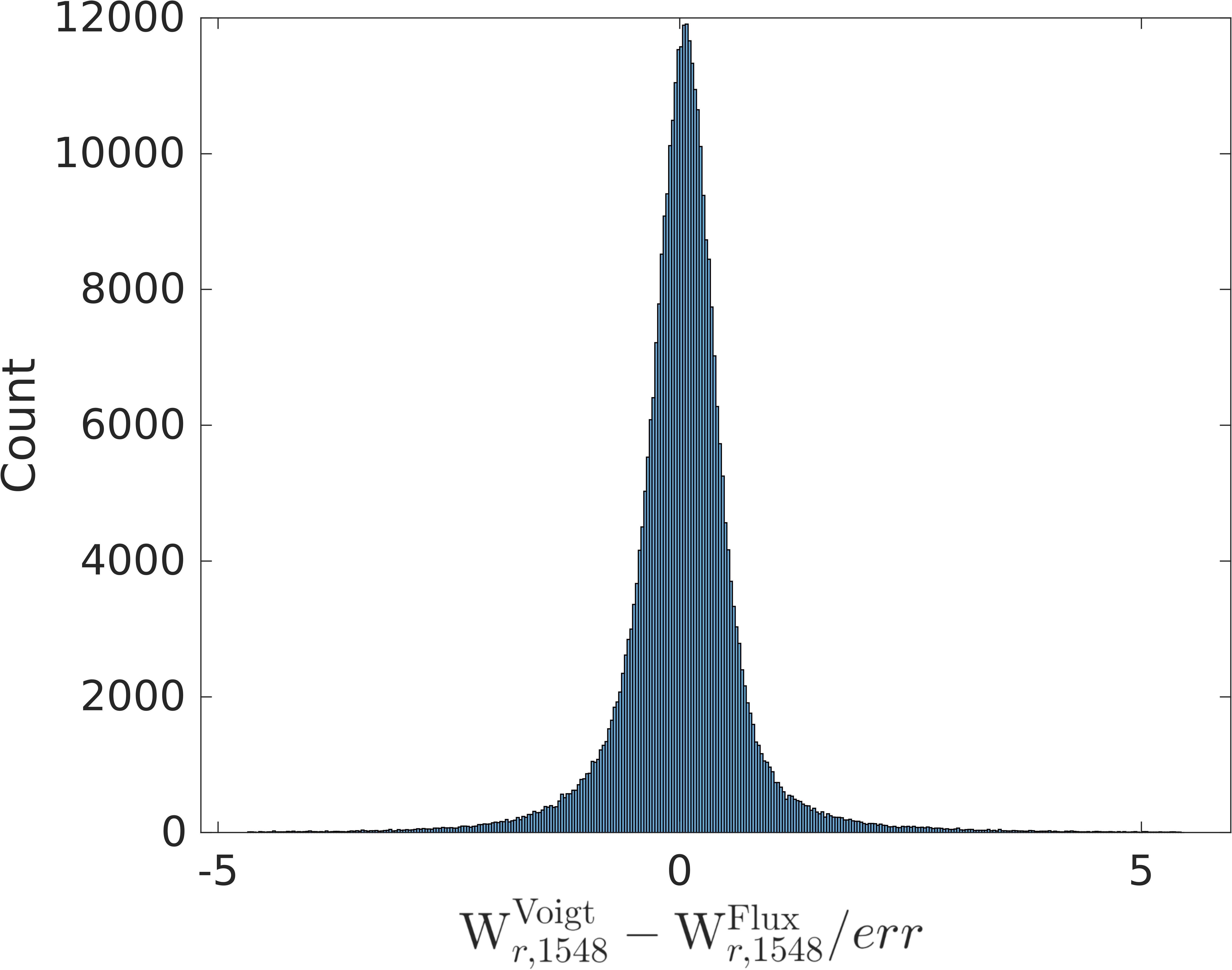}
  \caption{The difference between the two rest equivalent width estimates for the 1548~\AA~line explained in the text. These are using the maximum \textit{a posteriori} model parameters and integrating the flux around the detected \civ absorber. Differences are normalised by the expected error from the model parameter posteriors, and show the expected Gaussian distribution.
  }
  \label{fig:errDiffREW}
\end{figure}

As a consistency check, we compared the rest equivalent width from the maximum \textit{a posteriori} values of our model fit, using Equation \ref{eq:rewgp}, to the rest equivalent width from integrating the flux around the absorber, as in the validation phase. Figure \ref{fig:errDiffREW} shows the difference between the two rest equivalent width estimates, normalised by the error estimate. Figure \ref{fig:errDiffREW} shows a symmetric unit Gaussian distribution centered at zero, demonstrating that our model parameters are both approximately unbiased and have well-calibrated error estimates.


In the validation phase (see Section
\ref{sec:ValidREW}), we calculated the rest equivalent width by integrating the flux around the absorber, in order to compare to the rest equivalent widths from the PM catalogue. However, we prefer to estimate rest equivalent widths for our SDSS DR12 catalogue directly from our maximum \textit{a posteriori} model parameters, as these are less sensitive to noisy pixels in the integration range.

\begin{table}
  \caption{Table of probabilities P($\model_S$): first column shows the number of single line absorbers. The 2nd through 4th
   columns   show the number of single~absorbers with probabilities > 65\%, 85\%, and 95\% respectively. } 
  \label{tab:p_L1}
    \begin{tabular}{|l|l|l|l|}
    \hline
    \civ &	${\rm P}(\model_S)>0.65$ &	${\rm P}(\model_S)>0.85$&	${\rm P}(\model_S)>0.95$\\ \hline
    0 &	155905 (84.0\%) &	162533 (87.6\%) &	166675 (89.9\%)\\
    1	& 25441	(13.7\%) & 19159	(10.3\%) & 15366 (8.3\%)\\
    2	& 3210 (1.7\%) &	2914 (1.6\%)	& 2626 (1.4\%)\\
    3	& 675	(0.4\%) & 637 (0.3\%) &	583 (0.3\%)\\
    4 &	161 (0.09\%) &	152 (0.08\%) &	147 (0.08\%)\\
    5	& 29 (0.02\%)	& 26	(0.01\%) & 24 (0.01\%)\\
    6 &	4	(0.00\%) & 4 (0.00\%) &	4 (0.00\%)\\
      \end{tabular}
\end{table}

Table \ref{tab:p_L1} shows the number of candidate absorbers for the single line absorber model in SDSS DR12. Note that our training set does not label these absorbers and so we have not validated the potential detections.

%



\end{document}